\documentclass[11pt]{article}

\pdfoutput=1

\usepackage[T1]{fontenc}
\usepackage[latin9]{inputenc}
\usepackage[a4paper]{geometry}
\usepackage[active]{srcltx}
\usepackage{amsmath}
\usepackage{amssymb}
\usepackage{esint}
\usepackage{ulem}
\usepackage{cancel}
\makeatletter

\usepackage{textcomp}

\pdfoutput=1 

\usepackage{jheppub}

\usepackage{etoolbox}
    
    \patchcmd{\maketitle}{\@fpheader}{}{}{}

\usepackage{amsfonts}

\usepackage{bbm}

\newcommand{\be}{\begin{equation}}
\newcommand{\ee}{\end{equation}}
\newcommand{\bea}{\begin{eqnarray}}
\newcommand{\eea}{\end{eqnarray}}

\title{\boldmath Asymptotic structure of the Rarita-Schwinger theory in four spacetime dimensions at spatial infinity}

\author[a]{Oscar Fuentealba,}
\author[a,b]{Marc Henneaux,}
\author[a]{Sucheta Majumdar,}
\author[a]{Javier Matulich,}
\author[a]{and Turmoli Neogi}

\affiliation[a]{Universit\'e Libre de Bruxelles and International Solvay Institutes, ULB-Campus Plaine CP231, B-1050 Brussels, Belgium}
\affiliation[b]{Coll\`ege de France, 11 place Marcelin Berthelot, 75005 Paris, France}

\emailAdd{ofuentea@ulb.ac.be}
\emailAdd{henneaux@ulb.ac.be}
\emailAdd{sucheta.majumdar@ulb.ac.be}
\emailAdd{jmatulic@ulb.ac.be}
\emailAdd{turmoli.neogi@ulb.ac.be}

\preprint{}

\abstract{We investigate the asymptotic structure of the free Rarita-Schwinger theory 
in four spacetime dimensions at spatial infinity in the Hamiltonian formalism. We impose boundary conditions 
for the spin-3/2 field that are invariant under an infinite-dimensional (abelian) algebra of non-trivial asymptotic fermionic  symmetries. The compatibility of this set of boundary conditions with the invariance 
of the theory under Lorentz boosts requires the introduction of boundary degrees of freedom in 
the Hamiltonian action, along the lines of electromagnetism. These boundary degrees of freedom modify the symplectic structure by a surface contribution appearing in addition to the standard bulk piece. 
The Poincar\'e transformations have then well-defined (integrable, finite) canonical generators.  Moreover,
improper fermionic gauge symmetries, which  are also well-defined canonical transformations, are further enlarged and turn out to be parametrized 
by  two independent angle-dependent spinor functions at infinity, which lead to an infinite-dimensional fermionic algebra endowed 
with a central charge. We extend next the analysis to the supersymmetric spin-$(1,3/2)$ and spin-$(2,3/2)$ multiplets.  First, we present the canonical realization 
of the super-Poincar\'e algebra on the spin-$(1,3/2)$ multiplet, which is shown to be consistently enhanced by the infinite-dimensional abelian algebra of angle-dependent bosonic and fermionic  improper gauge symmetries  associated with the electromagnetic and the Rarita-Schwinger fields, respectively.
 A similar analysis of the spin-$(2,3/2)$ multiplet is then carried out to obtain the canonical realization 
of the super-Poincar\'e algebra, consistently enhanced by the abelian improper bosonic gauge transformations of the spin-$2$ field (BMS supertranslations) and the abelian improper fermionic gauge transformations of the spin-$3/2$ field. }

\makeatother

\begin{document}
\maketitle \flushbottom

 \newpage{}

\section{Introduction}

The asymptotic structure of gravity  is an extremely rich subject, which has attracted a revived interest in the last years  \cite{Strominger:2017zoo}.

One of the lessons that has been learned from the study of the behaviour of the gravitational field at infinity is that the algebra of asymptotic symmetries can be much larger than the algebra of background isometries.  The first instance where this phenomenon was observed was four-dimensional Einstein gravity with vanishing cosmological constant, where studies at null infinity revealed that the asymptotic symmetries formed the infinite-dimensional Bondi-van der Burg-Metzner-Sachs (BMS) algebra \cite{Bondi:1962px,Sachs:1962wk,Sachs:1962zza,Penrose:1962ij,Madler:2016xju,Alessio:2017lps,Ashtekar:2018lor}, which can even be further extended to include super-rotations \cite{Banks:2003vp,Barnich:2009se,Barnich:2010eb,Barnich:2011ct}.  Another much studied example is anti-de Sitter gravity in three spacetime dimensions, where the asymptotic symmetry algebra is given, for standard AdS boundary conditions,  by two copies of the infinite-dimensional  Virasoro algebra \cite{Brown:1986nw}.  In both cases, the asymptotic symmetry algebra contains the algebra of background isometries (the Poincar\'e algebra or the anti-de Sitter algebra, respectively), but is strictly bigger than it.   Furthermore, a non trivial central charge with interesting physical significance may appear in the Poisson bracket algebra of the generators of the asymptotic symmetries - which is necessarily trivial for the subalgebra of background isometries - , as it is the case for three-dimensional AdS gravity \cite{Brown:1986nw}. 

When one considers the supersymmetric versions of these two theories, one finds in both cases a graded extension of the respective infinite-dimensional symmetry algebras.  This was shown to be the case for supergravity in four dimensions  in \cite{Awada:1985by},  where a graded extension of the BMS algebra was exhibited to be a symmetry of the theory. While the analysis of  \cite{Awada:1985by} was performed at null infinity, it was later confirmed in \cite{Henneaux:2020ekh} that the same superalgebra  emerges at spatial infinity.  In the three-dimensional  context,  the infinite-dimensional asymptotic symmetry superalgebras of anti-de Sitter supergravities were worked out in \cite{Banados:1998pi,Henneaux:1999ib} and identified with (nonlinear) extended super-conformal algebras.

The main motivation of our paper, which deals with asymptotically flat spaces in four spacetime dimensions,  arises from the following puzzle.  A notable feature of the super-symmetric extension of the BMS algebra of \cite{Awada:1985by} is that it contains only a finite number of fermionic generators.  However, group theoretical arguments indicate that extensions of the BMS algebra by an infinite number of fermionic generators exist \cite{Awada:1985by} and indeed, such super-BMS algebras appeared more recently in the analysis of supergravity at null infinity \cite{Dumitrescu:2015fej,Lysov:2015jrs,Avery:2015iix,Fotopoulos:2020bqj,Narayanan:2020amh}.  The question we would like to explore is whether one can develop a consistent Hamiltonian formulation of this infinite-dimensional fermionic extension.  Such a step would shed important light on the quantum formulation of the theory. 

Indications that it should be possible to implement consistently such an extension comes actually from asymptotically flat supergravity in three dimensions, where a super-BMS$_3$ algebra possessing this feature was uncovered \cite{Barnich:2014cwa} (see also \cite{Banerjee:2016nio,Lodato:2016alv,Banerjee:2017gzj,Basu:2017aqn,Fuentealba:2017fck,Caroca:2018obf,Caroca:2019dds} for extended supergravity).

By consistent Hamiltonian formulation of the symmetry, we specifically mean the following: can one impose boundary conditions on the phase space variables of supergravity so that the super-BMS transformations have a well-defined moment map?  

The phase space variables of supergravity are fields defined on a Cauchy hypersurface (or, which is sufficient for our purposes, ``asymptotically Cauchy hypersurface''), providing thereby complete initial data out of which the future (Cauchy) development, including the behaviour of the fields near null infinity, can in principle be derived. The boundary conditions are given at large distances on these Cauchy hypersurfaces, i.e., at spatial infinity.  

A first requirement to be imposed is that the boundary conditions make the action, and in particular, its kinetic term finite.  This guarantees the existence of a well-defined, finite, symplectic $2$-form leading to the standard Hamiltonian structure.  A second requirement is that the symmetries under investigation - here, super-BMS transformations - not only preserve the boundary conditions, but also the symplectic form and hence can be associated with a well-defined canonical generator.

This program was successfully achieved for the BMS group of general relativity in \cite{Henneaux:2018cst,Henneaux:2018hdj,Henneaux:2019yax}, where boundary conditions implementing the above requirements were given.  In fact, two inequivalent sets of boundary conditions consistent with the BMS symmetry were devised.   The development of initial data satisfying either of the boundary conditions at spatial infinity was investigated near null infinity, confirming incidentally as a by-product that the existence of a smooth null infinity resulting from regular initial data on a Cauchy hypersurface is far from granted - as forcefully stressed and vigorously studied in \cite{Friedrich2004,ValienteKroon:2002fa} and references therein.  A similar analysis was performed in electromagnetism for the angle-dependent $u(1)$ transformations  \cite{Henneaux:2018gfi}.

The question, then, is whether one can go beyond   the supergravity analysis of \cite{Henneaux:2020ekh}  by relaxing the boundary conditions given there in such a way that an infinite-dimensional fermionic extension (and not just a finite-dimensional one) of the BMS symmetry can be realized as asymptotic symmetry, with well-defined canonical generators. 

The problem is rather intricate because it involves on the bosonic side a generalized form of parity conditions \cite{Regge:1974zd}, \cite{Henneaux:2018cst,Henneaux:2018hdj,Henneaux:2019yax}.   We shall therefore first consider the simpler linearized models, which contain in the gravitational case already a good wealth of information on the interacting case \cite{Fuentealba:2020ghw} (but see \cite{Tanzi:2020fmt} for the Yang-Mills theory).  Free supersymmetric models are the scope of this paper.  In a subsequent work we shall come back to the full interacting theories. 

We first show  that the free Rarita-Schwinger field on a Minkowski background admits, with suitable  boundary conditions that are verified to be Poincar\'e invariant, an infinite number of non-trivial fermionic improper \cite{Benguria:1976in}
gauge symmetries parametrized by two independent functions of the angles. These form an abelian algebra with a non trivial central charge.   The construction needs the introduction  of surface degrees of freedom, which we describe.  These are analogous to the surface degrees of freedom needed for a consistent Lorentz-invariant description of electromagnetism \cite{Henneaux:2018gfi} (see also \cite{Gervais:1976ec}) and are somewhat reminiscent of the edge modes of \cite{Freidel:2020xyx}.

We then consider the free supersymmetric $(1, \frac32)$ and $(2, \frac32)$ multiplets and show that rigid supersymmetry is compatible with our boundary conditions, leading to an infinite set of fermionic symmetries (standard rigid supersymmetry and improper fermionic gauge symmetries).  The algebra of the charges is computed and the Lorentz transformation properties of the fermionic parameters are in particular written.

Our paper is organized as follows.  In Section {\bf \ref{sec:StandardRS}}, we recall the results  of   \cite{Henneaux:2020ekh} in the context of the free Rarita-Schwinger field and stress that with the boundary conditions adopted in this paper, there is no non trivial fermionic asymptotic symmetry.  We next analyze (Section {\bf \ref{sec:NewBCRS}}) the free Rarita-Schwinger theory with softened boundary conditions involving an improper gauge term.  We show that these conditions are compatible with Poincar\'e invariance -- specifically, the boosts -- if one add a surface field with specific Lorentz transformations.  This surface field enters the symplectic form through a surface integral at infinity, much in the same way as what was shown for electromagnetism in \cite{Henneaux:2018gfi}.  We then work out the Poincar\'e generators, which are  all well-defined (integrable and finite), as well as the generators of the fermionic improper gauge symmetries, which turn out to form an abelian superalgebra parametrized by two spinor functions of the angles, with a non trivial central extension. In Section {\bf \ref{sec:Spin132}}, we extend the analysis to the spin-$(1,3/2)$ multiplet and show compatibility with supersymmetry. This requires to define the supersymmetry transformations by including in them a field-dependent $u(1)$ gauge transformation, in order to leave the (pre-)symplectic form invariant.  In Section {\bf \ref{sec:Spin232}}, we achieve the same task for the spin-$(2,3/2)$ multiplet. We show that the algebra of the rigid symmetries and the improper gauge symmetries form a graded extension of the BMS algebra possessing an infinite number of fermionic generators - the subalgebra of rigid symmetries and bosonic improper gauge symmetries being isomorphic to the super-BMS algebra of \cite{Awada:1985by,Henneaux:2020ekh}.  Finally, Section {\bf \ref{sec:Conclusions}} gives some concluding remarks. Two appendices provide conventions and some more technical material.

\section{Formulation of the Rarita-Schwinger theory with standard asymptotic conditions}
\label{sec:StandardRS}

We consider the Rarita-Schwinger field $\psi_\mu$ in four-dimensional Minkowski space. In the canonical formalism, the dynamical variables are its spatial components $\psi_k$ and the action is given in flat coordinates by
\cite{Deser:1977ur}
\begin{equation}
S=-\frac{i}{2}\int d^{4}x\,\overline{\psi}_{\mu}\gamma^{\mu\nu\rho}\partial_{\nu}\psi_{\rho}+\text{"boundary terms"}=\int dt\left(K-H-i\int d^{3}x\psi_{0}^{T}\mathcal{S}\right)\,.\label{eq:Action1}
\end{equation}
The kinetic term in the action reads explicitly\footnote{We choose the convention where Dirac conjugate for Majorana spinors
reads $\overline{\psi}_{\mu}=\psi_{\mu}^{T}\gamma_{0}$ with all $\gamma$-matrices
real, $\gamma_{0}$ being antisymmetric and $\gamma_{i}$ symmetric. }
\begin{equation}
K=\frac{i}{2}\int d^{3}x\,\psi_{k}^{T}\gamma^{km}\partial_{t}\psi_{m}+B\,,\label{eq:Kinetic-Term}
\end{equation}
where $B$ is a surface term chosen so that the action is finite. Its explicit form depends on the boundary conditions on the spin-$\frac32$ field.  The Hamiltonian is given by
\begin{equation}
H=\frac{i}{2}\int d^{3}x\,\psi_{k}^{T}\gamma_{0}\gamma^{kmn}\partial_{m}\psi_{n}\,.
\end{equation}
The constraint $\mathcal{S}$ enforced by varying the Lagrange multiplier
$\psi_{0}$ is
\begin{equation}
\mathcal{S}=\gamma^{mn}\partial_{m}\psi_{n}\approx0\,.\label{eq:ContraintS}
\end{equation}

The standard asymptotic conditions on $\psi_k$ are
\be
\psi_k = \mathcal{O}\left(\frac{1}{r^2} \right).
\ee
With these boundary conditions, the boundary term $B$ can be taken to vanish and the kinetic term in the action is finite.  It implies the bracket relation
\be
\{\psi_{m}^{\alpha}(x),\psi_{n}^{\beta}(\bar{x})\}=\frac{i}{2}(\gamma_{n}\gamma_{m})^{\alpha\beta}\delta^{(3)}(x-\bar{x})
\ee
showing that $\psi_k$ is somehow canonically self-conjugate.

The Lagrange multiplier $\psi_0$ describes the gauge transformation involved in the time evolution of the system and is arbitrary.  We shall impose
\be
\psi_0 = \mathcal{O} \left(\frac{1}{r^2} \right)
\ee
so that the motion ``involves no gauge transformation at infinity''.  This is a convenient asymptotic gauge fixing condition, which clearly makes the constraint term in the action finite. 

The Hamiltonian action \eqref{eq:Action1} is invariant under infinitesimal Poincar\'e transformations obtained by taking the Poisson bracket of the fields with the Poincar\'e generators 
\be
\int d^{3}x(\xi\mathcal{E}+\xi^{m}\mathcal{P}_{m}) \, ,
\ee
where
\bea
&& \mathcal{E}  =\frac{i}{2}\Big[\psi_{k}^{T}\gamma^{kmn}\gamma_{0}\partial_{m}\psi_{n}+\partial_{k}(\psi_{j}^{T}\gamma^{j}\gamma_{0}\psi^{k})\Big]\,,\\
&& \mathcal{P}_{k}  =-i\Big[\frac{1}{2}\partial_{m}(\psi_{n}^{T}\gamma^{mn}\psi_{k})+\frac{1}{2}\psi_{m}^{T}\gamma^{mn}\partial_{k}\psi_{n}+\frac{1}{8}\partial_{p}(\psi_{m}^{T}\gamma^{mn}\gamma_{k}^{\,\,\,p}\psi_{n})\Big]\,.
\eea
Here,
\begin{align}
\xi & =b_{i}x^{i}+a^{\perp}\,,\\
\xi^{i} & =b_{\,\,\,j}^{i}x^{j}+a^{i}\,,
\end{align}
are the components of the vector fields that parametrize the Poincar\'e transformations.
The arbitrary constants $b^{i}$ and $b_{ij}=-b_{ji}$ parametrize
the boosts and spatial rotations, respectively. The constants $a^{\perp}$
and $a^{i}$ stand for standard translations (the term $b^i x^0$ in $\xi^i$ can be absorbed in $a^i$ at any fixed time).
Note that the integral giving the Poincar\'e generators converges at infinity even for boosts and spatial rotations since in that case the integrand behaves as $\sim \frac{1}{r^4}$. 

One gets
\bea \delta_{\xi}\psi_{p}&=&\{\psi_{p},\int d^{3}x(\xi\mathcal{E}+\xi^{m}\mathcal{P}_{m})\} \nonumber  \\
&=& 
\xi\gamma_{p}^{\,\,\,jk}\gamma_{0}\partial_{j}\psi_{k}+\frac{1}{2}\partial_{j}\xi\gamma^{j}\gamma_{0}\psi_{p}-\frac{1}{2}\xi\gamma_{p}\gamma_{0}\mathcal{S}+\mathcal{L}_{\xi}\psi_{p}\,,\label{eq:Poincare-Transf0}
\eea
where the Lie derivative of $\psi_{k}$ is given by
\begin{equation}
\mathcal{L}_{\xi}\psi_{k}=\xi^{m}\partial_{m}\psi_{k}+\partial_{k}\xi^{m}\psi_{m}+\frac{1}{4}\partial_{m}\xi_{n}\gamma^{mn}\psi_{k}\,.
\end{equation}
By construction, this transformation leaves the symplectic form invariant since it is a canonical transformation.  Furthermore, it is direct to check that it  leaves the constraint invariant, $\delta_\xi {\mathcal S} \approx 0$.

The Hamiltonian action \eqref{eq:Action1} is also invariant under fermionic
gauge transformations of the form
\begin{equation}
\delta_{\eta}\psi_{k}=\partial_{k}\eta\,, \label{eq:Gauge0}
\end{equation}
where $\partial_{k}\eta$ must be of order $r^{-2}$ in order to preserve the asymptotic decay of the spin-$\frac32$ field, i.e., 
\be
\eta = \eta_0 + \mathcal{O}\left(\frac{1}{r} \right)
\ee
the constant term $\eta_0$ being ineffective in (\ref{eq:Gauge0}). This transformation is again canonical, being generated by 
\be
Q[\eta] = i \int d^{3}x\,(\eta - \eta_0)^T \mathcal{S},
\ee
an expression that can easily be verified again to converge at infinity.  One has indeed
\be
\delta_{\eta}\psi_{k}=\{\psi_{k},Q[\eta]\}\, .
\ee

The generator $Q[\eta]$ can be rewritten as
\be
Q[\eta] =i \int d^{3}x\,{\eta}^T  \mathcal{S} {-} i \eta_0^T \oint d^2 S_k  \gamma^{km} \psi_m \, ,
\ee
since $- \int d^{3}x \,   \mathcal{S} +  \oint d^2 S_k  \gamma^{km} \psi_m =0$ (Stokes theorem).  When the constraints hold, the generator $Q[\eta]$ vanishes and so, the gauge transformations are all proper \cite{Benguria:1976in}.  This is also true  for the zero mode $\eta_0$, as it should, since $\eta_0$ drops from the gauge transformations.   In the presence of couplings, of course, the constraint-generator $\mathcal{S}$  is not linear anymore.  There is a source contribution that makes the surface integral multiplying $\eta_0$ non-zero. But even in that case, there are only four improper fermionic gauge transformations parametrized by the zero mode $\eta_0$ because the generators of the other fermionic gauge symmetries are (weakly) equal to zero \cite{Henneaux:2020ekh}. 

The absence of improper gauge transformations in the free theory is due to our choice of  $\mathcal{O}(r^{-2})$ fall-off of the Rarita-Schwinger field. We shall now discuss less restrictive conditions that lead to a much more interesting asymptotic symmetry structure. 

\section{New asymptotic conditions for the Rarita-Schwinger field}
\label{sec:NewBCRS}

\subsection{Formulation of the new conditions}

The less restrictive boundary conditions developed in this paper allow for a $\mathcal{O}(r^{-1})$ term in the gravitino field $\psi_k$.  However, if one were to consider an arbitrary $\mathcal{O}(r^{-1})$ behaviour, one would encounter divergences in the symplectic structure and in the Lorentz generators that are difficult to tame.  For that reason, the allowed $\mathcal{O}(r^{-1})$ in $\psi_k$ will be assumed to take the specific form of a gradient. 
We therefore adopt as asymptotic conditions
\begin{align}
\psi_{k} & =\partial_{k}\chi+\mu_{k}\,,
\end{align}
where 
\begin{equation}
\chi=\overline{\chi}+\mathcal{O}\left(\frac{1}{r}\right)\, ,
\end{equation}
and 
\begin{equation}
\mu_{k}=\frac{\bar{\mu}_{k}}{r^{2}}+\mathcal{O}\left(\frac{1}{r^{3}}\right)\, ,
\end{equation}
where the coefficients of the various powers of $r$ are functions of the angles. 
We call $\psi_k$ the ``complete gravitino field''  and refer to $\mu_{k}$ as the ``bulk field'' and $\overline{\chi}$ as the ``surface field''. 
The bulk field $\mu_k$ depends on
all three spatial coordinates $x^{k}$, while $\overline{\chi}$ depends
only on the angles of the $2$-sphere ``at infinity''.  It is convenient to extend, as we have done, the surface field into a bulk field $\chi$.  There is clearly some ambiguity in the process since higher powers in $r^{-1}$ in $\chi$ can be absorbed into $\mu_k$, but this will not be a problem (this redundancy will appear as a proper gauge symmetry). 

The inclusion of the surface field $\overline{\chi}$ in the boundary condition is the new feature
with respect to the previous treatment of \cite{Henneaux:2020ekh}.  One might argue at this point that because the new term $\partial_k \chi$ in the asymptotic behaviour of $\psi_k$ is a gradient, it should be irrelevant since it takes the form of a gauge transformation.  However, only proper gauge transformations correspond to redundancies in the description of the system \cite{Benguria:1976in}.  Improper gauge transformations, which have non-vanishing generators even on-shell, do change the physical state of the system.  The added gradient term $\partial_k \overline{\chi}$ does turn out to be an improper gauge transformation.   A similar extension of the boundary conditions by an improper gauge term that is the leading term in the asymptotic expansion of the field was considered earlier in \cite{Henneaux:2019yqq}.

The boundary conditions on $\mu_k$ will actually need to be strengthened in order for the kinetic term in the action to be finite.  Anticipating what we shall find below, we already impose the needed extra condition, which expresses that the 
constraint function $\mathcal{S}$ should decay one
power of $r^{-1}$ faster than what generically follows from the boundary
conditions on the fields, i.e., it should decay as $r^{-4}$. Thus we impose
\begin{equation}
\mathcal{S}=\mathcal{O}\left(\frac{1}{r^{4}}\right)\,,
\end{equation}
which is very reminiscent of the fall-off conditions on the constraints
in the case of electromagnetism \cite{Henneaux:2018gfi} and gravity
\cite{Henneaux:2018cst,Henneaux:2019yax}.

The new asymptotic conditions on the gravitino field are easily checked to be
preserved under Poincar\'e transformations, which we take to coincide with  (\ref{eq:Poincare-Transf0}),
\be
 \delta_{\xi}\psi_{p} =
\xi\gamma_{p}^{\,\,\,jk}\gamma_{0}\partial_{j}\psi_{k}+\frac{1}{2}\partial_{j}\xi\gamma^{j}\gamma_{0}\psi_{p}-{\frac{1}{2}\xi\gamma_{p}\gamma_{0}\mathcal{S}} + \mathcal L_\xi \psi_p  \,,\label{eq:Poincare-Transf}
\ee
leading to
\begin{align}
\delta_{\xi}\chi & =-\xi \gamma_0 \gamma^m \partial_m \chi + \frac{1}{2}\partial_{j}\xi\gamma^{j}\gamma_{0}\chi  +\mathcal{L}_{\xi}\chi \, ,  \label{eq:PoincareChi0}\\
\delta_{\xi}\mu_{p} & = \xi\gamma_{p}^{\,\,\,jk}\gamma_{0}\partial_{j}\mu_{k}+\frac{1}{2}\partial_{j}\xi\gamma^{j}\gamma_{0}\mu_{p}{-\frac{1}{2}\xi\gamma_{p}\gamma_{0}\gamma^{jk}\partial_{j}\mu_{k}} +\mathcal{L}_{\xi}\mu_p \, . \label{eq:PoincareMu0}
\end{align}
Since the split of $\psi_p$ into $\partial_p \chi$ and $\mu_p$ involves ambiguities, there is also some ambiguity in the expressions for $\delta_{\xi}\chi $ and $\delta_{\xi}\mu_{p} $, besides the usual gauge freedom. The above choice, such that $\delta_{\xi}\chi $ and $\delta_{\xi}\mu_{p} $ depend respectively only on $\chi $ and $\mu_{p} $, is particularly convenient.

The boundary conditions are also preserved by
gauge transformations of the form
\begin{align}
\delta_{\epsilon}\chi & =\epsilon\,,\qquad\delta_{\varepsilon}\mu_{m}=\partial_{m}\varepsilon\,,
\end{align}
spanned by two fermionic parameters, $\epsilon$ and $\varepsilon$,
whose asymptotic behaviors are given by
\begin{align}
\epsilon & =\overline{\epsilon}+\mathcal{O}\left(\frac{1}{r}\right)\,,\\
\varepsilon & =\frac{\overline{\varepsilon}}{r}+\mathcal{O}\left(\frac{1}{r^{2}}\right)\,.
\end{align}
As we shall see, these transformations involve both proper and improper gauge symmetries.

\subsection{Finiteness of the kinetic term}

That the addition of a gradient of order $\mathcal{O}(r^{-1})$ to the gravitino field is a non-trivial step can be immediately seen at two places.  First, the action - specifically the kinetic term - is superficially divergent.  Second, the integral giving the Lorentz generators also superficially diverges at infinity (logarithmically) since the integrand, which is not gauge invariant, behaves now as $r^{-3}$.  We examine these two problems successively, starting with a proper definition of the action.

If one decomposes  the Hamiltonian kinetic term \eqref{eq:Kinetic-Term}, one gets three kind of terms:
\begin{equation}
\frac{i}{2}\int d^{3}x\partial_{k}\chi^{T}\gamma^{km}\partial_{m}\dot{\chi}\,,
\end{equation}
which is formally linearly divergent;
\begin{equation}
\frac{i}{2}\int d^{3}x\partial_{k}\chi^{T}\gamma^{km}\dot{\mu}_{m}+\frac{i}{2}\int d^{3}x\mu_{k}^{T}\gamma^{km}\partial_{m}\dot{\chi}\,,
\end{equation}
which is formally logarithmically divergent; and
\begin{equation}
\frac{i}{2}\int d^{3}x\mu_{k}^{T}\gamma^{km}\dot{\mu}_{m}\,,
\end{equation}
which is finite.

We analyse in turn the two divergent pieces.
\begin{itemize}
\item The first term is equal to
\begin{equation}
\frac{i}{2}\int d^{3}x\partial_{k}\left(\chi^{T}\gamma^{km}\partial_{m}\dot{\chi}\right)\,,
\end{equation}
\end{itemize}
and so is a surface term that can be removed by subtracting it through
$B$ in \eqref{eq:Kinetic-Term}.
\begin{itemize}
\item The second term can be transformed into
\begin{equation}
i\int d^{3}x\dot{\chi}^{T}\gamma^{km}\partial_{k}\mu_{m}+i\oint d^{2}S_{k}\chi^{T}\gamma^{km}\dot{\mu}_{m}\,,\label{eq:2ndTerm}
\end{equation}
\end{itemize}
(using $a^{T}\gamma^{km}b=b^{T}\gamma^{km}a$ for anticommuting spinors)
up to a total time derivative that can again be absorbed in $B$.
The surface term in \eqref{eq:2ndTerm} is clearly finite while the
bulk term is also finite since we assume that the constraint  $\mathcal{S}$ decays as $\mathcal{O}(r^{-4})$. 

So, if we adjust properly the surface term $B$, we get for the kinetic
term
\begin{equation}
K=i\int d^{3}x\dot{\chi}^{T}\gamma^{km}\partial_{k}\mu_{m}+\frac{i}{2}\int d^{3}x\mu_{k}^{T}\gamma^{km}\dot{\mu}_{m}+i\oint d^{2}S_{k}\chi^{T}\gamma^{km}\dot{\mu}_{m}+B'\,,
\end{equation}
which is finite. The term $B'$ is the undetermined left-over piece
from $B$ that remains after the appropriate subtractions making the
action finite have been performed. Space covariance and convergence
requirements show that $B'$ should be proportional to $i\oint d^{2}S_{k}\chi^{T}\gamma^{km}\dot{\mu}_{m}$
and so henceforth we consider the most general kinetic term
\begin{equation}
K=i\int d^{3}x\dot{\chi}^{T}\gamma^{km}\partial_{k}\mu_{m}+\frac{i}{2}\int d^{3}x\mu_{k}^{T}\gamma^{km}\dot{\mu}_{m}+\frac{i}{2}\alpha\oint d^{2}S_{k}\chi^{T}\gamma^{km}\dot{\mu}_{m}\,,\label{eq:New-Kinetic}
\end{equation}
where $\alpha$ is assumed to be a non-zero arbitrary constant. Note
that for $\alpha=2$ one has that $B'=0$. We also note that the fields $\chi$ and $\mu_{k}$
enter separately in the action and not only through the sum $\psi_p$, specifically in the kinetic term.

\subsection{Equations of motion}

We now verify that the action has a true extremum with our boundary conditions, and not just an extremum up to surface terms. The action for the generalized boundary conditions reads as above
\begin{equation}
S=\int dt\Big[K-i\int d^{3}x\Big(\frac{1}{2}\mu_{k}^{T}\gamma_{0}\gamma^{kmn}\partial_{m}\mu_{n}+\psi_{0}^{T}\mathcal{S}\Big)\Big]\,,\label{eq:OldAction}
\end{equation}
where the kinetic term $K$ is now given by
\begin{equation}
K=i\int d^{3}x\big(\dot{\chi}^{T}\gamma^{mn}\partial_{m}\mu_{n}+\frac{1}{2}\mu_{m}^{T}\gamma^{mn}\dot{\mu}_{n}\big)+\frac{i}{2}\alpha\oint d^{2}S_{m}\chi^{T}\gamma^{mn}\dot{\mu}_{n}\,.
\end{equation}
We shall impose the same convenient asymptotic condition
\be
\psi_0 = \mathcal{O} \left(\frac{1}{r^2} \right)
\ee
on the Lagrange multiplier $\psi_0$ .

The variation of \eqref{eq:OldAction} with respect to $\psi_{0}$
leads to the fermionic constraint
\begin{equation}
\mathcal{S}=\gamma^{mn}\partial_{m}\mu_{n}=0\,.\label{eq:Constraint}
\end{equation}

The variation of the action with respect to $\mu_{m}$ turns out to
be
\begin{equation}
\delta S=i\int dt\Big\{ \int d^{3}x\delta\mu_{m}^{T}\Big[\gamma^{mn}(\dot{\mu}_{n}+\partial_{n}\dot{\chi}-\partial_n \psi_{0})-\gamma_{0}\gamma^{mnp}\partial_{n}\mu_{p}\Big]+\left(1-\frac{\alpha}{2}\right)\oint d^{2}S_{m}\delta \mu_{n}^{T}\gamma^{mn}\dot{\chi}\Big\}\,,\label{eq:deltaS}
\end{equation}
where the boundary conditions have been taken into account to get rid of
some boundary terms. If one imposes $\delta S = 0$,  the bulk contribution and the surface term must separately vanish. Vanishing of the bulk term reads
\begin{equation}
\gamma^{mn}(\dot{\mu}_{n}+\partial_{n}\dot{\chi}-\partial_{n}\psi_{0})-\gamma_{0}\gamma^{mnp}\partial_{n}\mu_{p}=0\,.\label{eq:SpatialRS}
\end{equation}
The leading order $1/r$ of \eqref{eq:SpatialRS} reduces to the condition
\begin{equation}
\gamma^{mk}\partial_{k}\dot{\overline{\chi}}=0\,,
\end{equation}
on the leading term $\overline{\chi}$ in the asymptotic expansion
of $\chi$, from which one infers
\begin{equation}
\partial_{t}(\partial_{k}\overline{\chi})=0\,. \label{eq:EOMchi}
\end{equation}
The zero mode of $\overline{\chi}$ is thus an arbitrary function of time,
but its higher spherical harmonic components are constant in time.  The subsequent orders in \eqref{eq:SpatialRS} are  just the standard dynamical Rarita-Schwinger equations for the  gravitino field. 
The boundary term in \eqref{eq:deltaS} then vanishes as a direct consequence
of  \eqref{eq:EOMchi} and the constraint
\eqref{eq:Constraint} (for the zero mode).

The variation of \eqref{eq:OldAction} with respect to $\chi$ yields
\begin{equation}
\delta S=\frac{i}{2}\alpha\oint d^{2}S_{m}\delta\chi^{T}\gamma^{mn}\dot{\mu}_{n}\,,
\end{equation}
modulo the constraint. The above implies the following boundary equation
of motion\footnote{Here, $x^{A}$ denotes the coordinates on the 2-sphere at infinity.
In spherical coordinates, for example, $x^{A}=(\theta,\varphi)$ labels
the two angles on the sphere.}
\begin{equation}
\gamma^{rA}\dot{\bar{\mu}}_{A}=0
\end{equation}
for the angular component of $\mu_k$.
Note that the next-to-leading order $(1/r^{2})$ of \eqref{eq:SpatialRS}
implies
\begin{equation}
\gamma^{rA}(\dot{\overline{\mu}}_{A}+\overline{D}_{A}\dot{\chi}^{(1)})=0\,,
\end{equation}
where $\chi^{(1)}$ is the coefficient of the $(1/r)$-term in the
expansion of $\chi$. This leads to the Dirac equation on the 2-sphere
for $\dot{\chi}^{(1)}$, from which one also gets that the zero mode
of $\chi^{(1)}$ is an arbitrary function of time, but its higher
spherical harmonic components are constant in time.

We can thus conclude that the action (\ref{eq:OldAction}) is a true extremum on the classical histories, which obey the Rarita-Schwinger equations of motion supplemented by (compatible) dynamical equations on the first terms in the development of $\chi$.

\subsection{Pre-symplectic form}

In order to discuss the canonical implementation of the Poincar\'e symmetry, we first need to understand the Poisson bracket structure of the theory.

The kinetic term \eqref{eq:New-Kinetic} in the action yields the
pre-symplectic form
\begin{equation}
\Omega=-i\int d^{3}xd_{V}\chi^{T}\gamma^{km}\partial_{k}d_{V}\mu_{m}+\frac{i}{2}\int d^{3}xd_{V}\mu_{k}^{T}\gamma^{km}d_{V}\mu_{m}+\frac{i}{2}\alpha\oint d^{2}S_{k}d_{V}\chi^{T}\gamma^{km}d_{V}\mu_{m}\,.\label{eq:Omega1}
\end{equation}
Note that since $\chi$ and $\mu_{m}$ are anticommuting, the one-forms
$d_{V}\chi$ and $d_{V}\mu_{m}$ are commuting. In particular, $d_{V}\mu_{k}^{T}\gamma^{km}d_{V}\mu_{m}$
is not zero. If the one-form $a$ is commuting, then $d_{V}(a\wedge b)=d_{V}a\wedge b+a\wedge d_{V}b$.

This closed 2-form $\Omega$ form is {\it degenerate}, i.e., there
exists a non-vanishing $Y$ such that $\iota_{Y}\Omega=0$. Our goal in this subsection is to determine all the null vectors $Y$ of $\Omega$.

The equation $\iota_{Y}\Omega=0$ reads
\begin{align}
0 & =-i\int d^{3}x\Upsilon^{T}\gamma^{km}\partial_{k}d_{V}\mu_{m}-i\int d^{3}xd_{V}\chi^{T}\gamma^{km}\partial_{k}\Sigma_{m}\nonumber \\
 & \quad+\frac{i}{2}\int d^{3}x\Sigma_{k}^{T}\gamma^{km}d_{V}\mu_{m}+\frac{i}{2}\int d^{3}xd_{V}\mu_{k}^{T}\gamma^{km}\Sigma_{m}\nonumber \\
 & \quad+\frac{i}{2}\alpha\oint d^{2}S_{k}\Upsilon^{T}\gamma^{km}d_{V}\mu_{m}+\frac{i}{2}\alpha\oint d^{2}S_{k}d_{V}\chi^{T}\gamma^{km}\Sigma_{m}\,,
\end{align}
with
\begin{equation}
\delta_{Y}\chi=\Upsilon\,,\qquad\delta_{Y}\mu_{m}=\Sigma_{m}\,.
\end{equation}

The terms involving $d_{V}\chi$ and $d_{V}\mu_{m}$ are independent
and must vanish separately. Thus, one must have
\begin{equation}
0=-i\int d^{3}x\Upsilon^{T}\gamma^{km}\partial_{k}d_{V}\mu_{m}+i\int d^{3}x\Sigma_{k}^{T}\gamma^{km}d_{V}\mu_{m}+\frac{i}{2}\alpha\oint d^{2}S_{k}\Upsilon^{T}\gamma^{km}d_{V}\mu_{m}\,,\label{eq:Eq1}
\end{equation}
and
\begin{equation}
0=-i\int d^{3}xd_{V}\chi^{T}\gamma^{km}\partial_{k}\Sigma_{m}+\frac{i}{2}\alpha\oint d^{2}S_{k}d_{V}\chi^{T}\gamma^{km}\Sigma_{m}\,.\label{eq:Eq2}
\end{equation}

The first line can be transformed into
\begin{equation} \label{Null-Variation-Pre-Symp}
0=i\int d^{3}x\left(\partial_{k}\Upsilon^{T}+\Sigma_{k}^{T}\right)\gamma^{km}d_{V}\mu_{m}-i\left(1-\frac{\alpha}{2}\right)\oint d^{2}S_{k}\Upsilon^{T}\gamma^{km}d_{V}\mu_{m}\,,
\end{equation}
which will be zero for arbitrary $d_{V}\mu_{m}$ satisfying the boundary
conditions if the bulk and surface terms separately vanish. Vanishing
of the bulk term imposes $\Sigma_{k}=-\partial_{k}\Upsilon$ from which
it follows in particular that the leading term $\overline{\Upsilon}$
in $\text{\ensuremath{\Upsilon}}$ is constant since $\Sigma_k = \mathcal{O}(r^{-2})$, i.e., $\Upsilon=C+\frac{\Upsilon^{(1)}(x^{A})}{r}+\mathcal{O}(r^{-2})$. {The constant $C$ is unrestricted if $\alpha = 2$ but must vanish for the surface term involving $d_{V}\mu_{m}$ to be zero when $\alpha \not=2$. }

Vanishing of the bulk term in the second equation \eqref{eq:Eq2}
is an immediate consequence of $\Sigma_{k}=-\partial_{k}\Upsilon$, and
vanishing of its surface term implies that $\Upsilon^{(1)}$ is also a
constant ($\alpha \not=0$), so that we get as zero vector fields $Y$ of $\Omega$
\begin{equation}
\Upsilon=C+\frac{C^{(1)}}{r}+\mathcal{O}(r^{-2})\,,\qquad\Sigma_{k}=-\partial_{k}\Upsilon\, {, \quad (\alpha = 2)}\label{eq:Null-Y}
\end{equation}
{(and $C=0$ if $\alpha \not=2$).}
When the pre-symplectic form is degenerate, the correspondence between phase space Hamiltonian vector fields and phase space functions is amended in an obvious way.  A phase space vector field $X$ is still called Hamiltonian if $\mathcal{L}_X \Omega = 0$, which is equivalent to $d_V (\iota_X \Omega) = 0$. This implies $\iota_X \Omega = - d_V F$ for some function $F$ that is defined up to a constant and that is necessarily constant on the null submanifolds generated by the null vectors $Y$, since $0 = \iota_Y d_V F = \mathcal{L}_Y  F $. Conversely, a phase space function $F$ defines a Hamiltonian vector field only if it is constant on the null submanifolds generated by the null vectors $Y$, and the corresponding Hamiltonian vector field is then defined only up to a combination of the $Y$'s.   A recent instance where such a phenomenon occurs in an asymptotic analysis was studied in \cite{Henneaux:2020nxi}.

One can get a true symplectic structure by factoring out the null leaves of the pre-symplectic form.  This can be done by imposing a gauge condition that freezes the possibility to move along the null leaves. One way to achieve this in our case is to impose that $\chi$ reduces to its first two terms in the asymptotic expansion, and that the coefficients $\overline \chi$ and $\chi^{(1)}$ have no zero mode.

\subsection{Canonical realization of the boosts - Need for a new surface field}

It is straightforward to check that the pre-symplectic form is invariant
under spatial rotations and translations, given by (\ref{eq:PoincareChi0}) and (\ref{eq:PoincareMu0}) with $\xi = a^\perp$ and $\xi^i = b^i_{\; \, j} x^j + a^i$, so that these transformations are canonical by construction with a well-defined generator.  The situation is more complicated for boosts,
for which
\begin{align}
\delta_{\xi}\chi & =-\xi \gamma_0 \gamma^m \partial_m \chi + \frac{1}{2}\partial_{j}\xi\gamma^{j}\gamma_{0}\chi   \, ,  \\
\delta_{\xi}\mu_{p} & = \xi\gamma_{p}^{\,\,\,jk}\gamma_{0}\partial_{j}\mu_{k}+\frac{1}{2}\partial_{j}\xi\gamma^{j}\gamma_{0}\mu_{p}{-\frac{1}{2}\xi\gamma_{p}\gamma_{0}\gamma^{jk}\partial_{j}\mu_{k}}  \, ,
\end{align} 
with
\be
\xi = b_i x^i, \qquad (\xi^k = 0) \, .
\ee

The pre-symplectic form changes (off-shell) by the following surface term
\begin{align}
d_{V}(\iota_{\xi}\Omega)&=\frac{i}{2}\oint d^{2}S_{k}\xi d_{V}\mu_{m}^{T}\gamma^{kmn}\gamma_{0}d_{V}\mu_{n} +\frac{i}{2}\alpha\oint d^{2}S_{k}\xi\partial_{m}d_{V}\chi^{T} \gamma_0 \gamma^k d_V \mu^m \, .
\end{align}
The first term vanishes once the boundary conditions are taken into
account, while the second term is finite. The variation of the symplectic
form then becomes
\begin{equation}
d_{V}(\iota_{\xi}\Omega)=+\frac{i}{2}\alpha\oint d^{2}S_{k}\xi\partial_{m}d_{V}\chi^{T} \gamma_0 \gamma^k d_V \mu^m     \, ,\label{eq:Boosts}
\end{equation}
and is not zero, even if one takes into account that the constraints hold asymptotically.

In order for the boosts to be canonical transformations, we must find a way to get an invariant pre-symplectic form
\begin{equation}
d_{V}(\iota_{\xi}\Omega)=0\Rightarrow\iota_{\xi}\Omega=-d_{V}P_{\xi}\,,
\end{equation}
which would allow us to define a canonical generator $P_{\xi}$ for boosts.  To that end, we extend the phase space by adding new boundary degrees of freedom, the variation of which compensates the non-vanishing of (\ref{eq:Boosts}).

\subsubsection*{Integrability of the boost generator}

The non-vanishing of the right hand side of equation \eqref{eq:Boosts}
is due to the presence of the field $\chi$. This problem does not
arise in the supergravity analysis in \cite{Henneaux:2020ekh}, where
the boost generator for the spin-$3/2$ field is immediately integrable,
as there is no $\chi$ field in that construction. However, keeping
the field $\chi$ is crucial in our case in order to obtain an infinite-dimensional
set of improper gauge symmetries at spatial infinity. In analogy with
the case of electromagnetism \cite{Henneaux:2018gfi}, we cure the
problem by introducing an additional boundary field in the action
principle to make the symplectic form invariant under boosts.

More specifically, we  consider a vector-spinor field $\overline{\rho}^{k}$ depending only
on the angles of the $2$-sphere ``at infinity'', which extends
into the bulk as
\begin{equation}
\rho^{k}=\frac{\overline{\rho}^{k}}{r^{2}}+\mathcal{O}\left(\frac{1}{r^{3}}\right)\,,
\end{equation}
 endowed with the following symplectic form
\begin{equation}
\Omega_{0}=-\frac{i}{2}\alpha\oint d^{2}S_{k}\,d_{V}\chi^{T}\gamma_{0}d_{V}\rho^{k}\,.\label{eq:Symp0}
\end{equation}
We postulate the following infinitesimal transformation law under
the Poincar\'e group,
\begin{equation}
\delta_{b,Y}\overline{\rho}^r= -\gamma^r \overline{D}_A (b \overline{\mu}^A) -\frac{1}{2} b \overline{\gamma}_A\overline{\mu}^A +b\gamma_0 \overline{\gamma}^A \overline{D}_A \overline{\rho}^r + \frac{3}{2}b \gamma^r \gamma_0 \overline{\rho}^r -\frac{1}{2} \partial_A b \overline{\gamma}^A \gamma_0 \overline{\rho}^r    +\mathcal{L}_{Y}\overline{\rho}^{r}  \, ,\label{eq:delta-psi0}
\end{equation}
and for the angular components $\delta_{b,Y} \overline{\rho}^A=0$. (See Appendix \ref{Spherical} for spherical coordinates). \\

The change of the symplectic form \eqref{eq:Symp0} is given by
\begin{align}
d_{V}(\iota_{\xi}\Omega_{0}) & =-\frac{i}{2}\alpha\oint d^{2}S_{k}\left(d_{V}\delta_{\xi}\chi^{T}\gamma_{0}d_{V}\rho^{k}+d_{V}\chi^{T}\gamma_{0}d_{V}\delta_{\xi}\rho^{k}\right)\,.
\end{align}
For the boosts, this expression becomes, up to a total derivative on the $2$-sphere, 
\begin{align}
d_{V}(\iota_{\xi}\Omega_{0}) & =-\frac{i}{2}\alpha\oint d^{2}S_{k}\xi\partial_{m}d_{V}\chi^{T} \gamma_0 \gamma^k d_V \mu^m   \,.
\end{align}
Therefore, the total change of the symplectic form under boosts vanishes
\begin{equation}
d_{V}(\iota_{\xi}\Omega)+d_{V}(\iota_{\xi}\Omega_{0})=0\,,
\end{equation}
which makes the boosts canonical  transformations.

The introduction of the field $\rho^{k}$ modifies the kinetic term
by a boundary contribution which one easily reads from $\Omega_0$, leading to
\begin{equation}
K=i\int d^{3}x\big(\dot{\chi}^{T}\gamma^{mn}\partial_{m}\mu_{n}+\frac{1}{2}\mu_{m}^{T}\gamma^{mn}\dot{\mu}_{n}\big)+\frac{i}{2}\alpha\oint d^{2}S_{m}\big(\chi^{T}\gamma^{mn}\dot{\mu}_{n}-\chi^{T}\gamma_{0}\dot{\rho}^{m}\big)\,.\label{eq:new-Kinetic}
\end{equation}
The action principle with the additional boundary field now reads
\begin{equation}
S=\int dt\Big[K-i\int d^{3}x\Big(\frac{1}{2}\mu_{k}^{T}\gamma_{0}\gamma^{kmn}\partial_{m}\mu_{n}+\psi_{0}^{T}\gamma^{mn}\partial_{m}\mu_{n}\Big)\Big]\,.\label{eq:TotalAction}
\end{equation}
It is easy to see that the action is invariant under arbitrary 
shifts of $\rho^{m}$,
\begin{equation}
\delta_{\sigma}\rho^{m}=\sigma^{m}\,,\qquad\delta_{\sigma}\chi=0\,,\qquad\delta_{\sigma}\mu_{m}=0\,,
\end{equation}
where the parameter $\sigma^{m}$ falls off as
\begin{equation}
\sigma_{m}=\frac{\overline{\sigma}_{m}}{r^{2}}+\mathcal{O}\left(\frac{1}{r^{3}}\right)\,.
\end{equation}
It is useful to make the orthogonal decomposition of  $\overline{\sigma}_m$ as
\be
\overline{\sigma}^m = \overline{\sigma} n^m + \overline{\theta}^m
\ee
where $n^m = \frac{\partial}{\partial r}$ is normal to the sphere and $\overline{\theta}_m n^m =0$.   Invariance of the action forces $\overline{\sigma}$ to be time-independent but imposes no condition on $\overline{\theta}^m$ because only the radial component $\overline{\rho}^r$ of $\overline{\rho}^m$ appears in the action.  The angular components of $\overline{\rho}^m$ are absent and hence pure gauge degrees of freedom that can be ``gauged away'' (in fact they are already ``away''!). The transformations parametrized by $\overline{\theta}_m$ are proper gauge transformations and there is only one degree of freedom in $\overline{\rho}^m$, namely, its radial component $\overline{\rho}^r$. 

{The change of variables $\rho^m \rightarrow \omega^m$ with
\be
\omega^m = \rho^m + \gamma_0 \gamma^{mn} \mu_n  \quad \Rightarrow \quad \overline{\omega}^m = \overline{\rho}^m + \gamma_0 \gamma^{mn} \overline{\mu}_n  \label{omega}
\ee
brings the kinetic term $K$ to the simpler form
\begin{equation}
K=i\int d^{3}x\big(\dot{\chi}^{T}\gamma^{mn}\partial_{m}\mu_{n}+\frac{1}{2}\mu_{m}^{T}\gamma^{mn}\dot{\mu}_{n}\big)-\frac{i}{2}\alpha\oint d^{2}S_{m}\chi^{T}\gamma_{0}\dot{\omega}^{m} \,, \label{eq:new-Kinetic22}
\end{equation}
which clearly involves only the radial component $\overline{\omega}^r$. Henceforth, we keep the formulations both in terms of $\rho^k$ and $\omega^k$, as they enlighten different aspects of the theory.}

\subsubsection*{Pre-symplectic structure}

Given the boundary modification of the action, the pre-symplectic form becomes
\begin{align}
\Omega & =-i\int d^{3}xd_{V}\chi^{T}\gamma^{km}\partial_{k}d_{V}\mu_{m}+\frac{i}{2}\int d^{3}xd_{V}\mu_{k}^{T}\gamma^{km}d_{V}\mu_{m}\nonumber \\
 & \quad+\frac{i}{2}\alpha\oint d^{2}S_{k}d_{V}\chi^{T}\gamma^{km}d_{V}\mu_{m}-\frac{i}{2}\alpha\oint d^{2}S_{k}\,d_{V}\chi^{T}\gamma_{0}d_{V}\rho^{k}\,.
\end{align}
{Note that in terms of $\omega^k$, the surface terms are combined into a single term:
\be
-\frac{i}{2}\alpha\oint d^{2}S_{k}\,d_{V}\chi^{T}\gamma_{0}d_{V}\omega^{k}\,.
\ee
}
The null vectors $Y$ are accordingly given by new expressions that are easily worked out. 
A direct computation shows that these are now (with $\delta_{Y}\chi=\Upsilon$, $\delta_{Y}\mu_{m}=\Sigma_{m}$ and $\delta_{Y} \rho^k = R^k$),
\begin{equation}
\Upsilon=\frac{C^{(1)}}{r}+\mathcal{O}(r^{-2})\,,\quad\Sigma_{k}=-\partial_{k}\Upsilon\,, \quad R^r =  \gamma^{rA} \partial_A \Upsilon    \, , \quad R^A =  y^A\, ,\label{eq:Null-Y223}
\end{equation}
where $y^A$ are arbitrary functions. In particular, the zero mode of $\overline{\chi}$ is not pure gauge anymore since the constant $C$ of (\ref{eq:Null-Y}) is now forced to be equal to zero in (\ref{eq:Null-Y223}) {even when $\alpha = 2$}.  This is because the zero mode of $\rho^r$ is present in the new action.

\subsubsection*{Equations of motion}

The boundary modification of the action leads also to a slight change in the boundary equations of motion.

Variation of the complete action \eqref{eq:TotalAction} with respect to $\rho^k$ yields $\dot{\overline{\chi}} = 0$. This boundary equation of motion is a consequence of the unchanged equations of motion obtained by varying $\mu_k$,  with the only additional information that the zero mode of $\overline{\chi}$ should also be constant in time (instead of being arbitrary). 

If one varies the complete action \eqref{eq:TotalAction} with respect to $\chi$, one gets 
\begin{equation}
\delta S=\frac{i}{2}\alpha\oint d^{2}S_{m}\delta\chi^{T}(\gamma^{mn}\dot{\mu}_{n}-\gamma_{0}\dot{\rho}^{m})\,,
\end{equation}
modulo the constraint. The boundary equation of motion is thus now given
by
\begin{equation}
\gamma^{rA}\dot{\bar{\mu}}_{A}-\gamma_{0}\dot{\bar{\rho}}^{r}=0\, \quad {\Leftrightarrow \quad \dot{\overline{\omega}}^r=0}\, ,
\end{equation}
an equation that can be viewed as fixing $\dot{\overline{\rho}}$ in terms of $\dot{\bar{\mu}}_{A}$.

\subsection{Poincar\'e generators}

The addition of the new boundary field renders all the Poincar\'e transformations
canonical, and consequently leads to well-defined generators through
its relation with the symplectic form
\begin{equation}
\iota_{X}\Omega=-d_{V}G\,.
\end{equation}

With the expression of the infinitesimal Poincar\'e transformations of the fields written above, 
\begin{align}
\delta_{\xi}\chi & =-\xi \gamma_0 \gamma^m \partial_m \chi + \frac{1}{2}\partial_{j}\xi\gamma^{j}\gamma_{0}\chi  +\mathcal{L}_{\xi}\chi \, ,  \label{eq:PoincareChi}\\
\delta_{\xi}\mu_{p} & = \xi\gamma_{p}^{\,\,\,jk}\gamma_{0}\partial_{j}\mu_{k}+\frac{1}{2}\partial_{j}\xi\gamma^{j}\gamma_{0}\mu_{p}{-\frac{1}{2}\xi\gamma_{p}\gamma_{0}\gamma^{jk}\partial_{j}\mu_{k}} +\mathcal{L}_{\xi}\mu_p \, . \label{eq:PoincareMu}  \\
\delta_{b,Y}\overline{\rho}^r &= -\gamma^r \overline{D}_A (b \overline{\mu}^A) -\frac{1}{2} b \overline{\gamma}_A\overline{\mu}^A +b\gamma_0 \overline{\gamma}^A \overline{D}_A \overline{\rho}^r + \frac{3}{2}b \gamma^r \gamma_0 \overline{\rho}^r -\frac{1}{2} \partial_A b \overline{\gamma}^A \gamma_0 \overline{\rho}^r    +\mathcal{L}_{Y}\overline{\rho}^{r}  \nonumber \\
&\Leftrightarrow\quad \delta_{b,Y} \overline{\omega}^r ~=~ b \gamma_0 \overline{\gamma}^A \overline{D}_A \overline{\omega}^r + \frac{3}{2} b \gamma^r \gamma_0 \overline{\omega}^r - \frac{1}{2} \partial_A b \overline{\gamma}^A \gamma_{0} \overline{\omega}^r + \mathcal{L}_Y \overline{\omega}^r
\end{align}
one finds that the Poincar\'e generators are explicitly given by
\begin{align}
P_{\text{\ensuremath{\xi},\ensuremath{\xi^{i}}}} & =\int d^{3}x\left(\xi\mathcal{H^{\text{RS}}}+\xi^{i}\mathcal{H}_{i}^{\text{RS}}\right)+\mathcal{B}_{\xi,\xi^{i}}^{\text{RS}}\,,\label{eq:Poincare-RS}\\
\mathcal{H^{\text{RS}}} & =\frac{i}{2}\Big[\mu_{k}^{T}\gamma^{kmn}\gamma_{0}\partial_{m}\mu_{n}+\partial_{k}(\mu_{j}^{T}\gamma^{j}\gamma_{0}\mu^{k})+\partial_{k}(\chi^{T}\gamma^{k}\gamma_{0}\gamma^{mn}\partial_{m}\mu_{n}) \nonumber \\
& \quad +2 \partial_k \chi^T \gamma_0 \gamma^k \gamma^{pq}\partial_p\mu_q           \Big]\, , \\
\mathcal{H}_{i}^{\text{RS}} & =-i\Big[\frac{1}{2}\partial_{m}(\mu_{n}^{T}\gamma^{mn}\mu_{k})+\frac{1}{2}\mu_{m}^{T}\gamma^{mn}\partial_{k}\mu_{n}+\frac{1}{8}\partial_{p}(\mu_{m}^{T}\gamma^{mn}\gamma_{k}^{\,\,\,p}\mu_{n})\nonumber \\
 & \quad+\partial_{k}\chi^{T}\gamma^{mn}\partial_{m}\mu_{n}+\frac{1}{4}\partial_{p}(\chi^{T}\gamma_{k}^{\,\,\,p}\gamma^{mn}\partial_{m}\mu_{n})\Big]\,,\\
\mathcal{B}_{\xi,\xi^{i}}^{\text{RS}} & = \frac{i}{2}\alpha  \oint d^2S_k \left( \delta_{\xi} \chi^T \gamma_0 \omega^k + \mathcal{L}_{\xi_i}\chi^T \gamma_0 \omega^k  \right)  \,.
\end{align}
where $\omega^k$ is defined in \eqref{omega}.

\vskip 0.1cm 
\subsection{A  {twofold} of infinite-dimensional fermionic improper gauge symmetries}
Similarly, the canonical generator of the proper and improper gauge symmetries reads
\begin{align}
G_{\epsilon,\varepsilon,\sigma^{k}}^{\text{RS}} & =i\int d^{3}x\left(\epsilon^{T}+\varepsilon^{T}\right)\gamma^{km}\partial_{k}\mu_{m}\nonumber \\
 & \quad+\frac{i}{2}\alpha\oint d^{2}S_{k}\left(\epsilon^{T}\gamma^{km}\mu_{m}-\chi^{T}\gamma^{km}\partial_{m}\varepsilon-\epsilon^{T}\gamma_{0}\rho^{k}+\chi^{T}\gamma_{0}\sigma^{k}\right)\,,\label{eq:Gfermion}
\end{align}
generating the infinitesimal transformation laws of the fields
\begin{align}
\delta_{\epsilon}\mathcal{\chi} & =\epsilon\,,\label{eq:GaugeChi}\\
\delta_{\varepsilon}\mu_{m} & =\partial_{m}\varepsilon\,,\label{eq:GaugeMu}\\
\delta_{\sigma_{m}}\rho^{m} & =\sigma^{m}\,.\label{eq:GaugeRho}
\end{align}

{One can rewrite the generator (\ref{eq:Gfermion}) in terms of the new variable $\omega^k$ as
\begin{align}
G_{\epsilon,\varepsilon,\sigma^{k}}^{\text{RS}} & \equiv G_{\epsilon,\varepsilon,\zeta^{k}}^{\text{RS}} \nonumber \\
& =i\int d^{3}x\left(\epsilon^{T}+\varepsilon^{T}\right)\gamma^{km}\partial_{k}\mu_{m}
  + \frac{i}{2}\alpha\oint d^{2}S_{k}\left(\chi^{T}\gamma_{0}\zeta^{k} - \epsilon^{T}\gamma_{0}\omega^{k}\right)\,,\label{eq:Gfermion22}
\end{align}
where we have set $\zeta^k = \sigma^k + \gamma_0 \gamma^{km} \partial_m \varepsilon$, so that the transformations of the field $\omega^k$ read
\be
\delta_{\zeta_{m}}\omega^{m}  =\zeta^{m}\,.\label{eq:GaugeOmega}
\ee}
The bulk term of the charge \eqref{eq:Gfermion22} is proportional to
the constraint. The surface integral to which the charge reduces on-shell
does not vanish if  the leading terms $\overline{\epsilon}$ {
and $\overline{\zeta}^{r}$} in the expansions
of $\epsilon$ {and $\zeta^{r}$} are not zero.  These are the improper gauge symmetries.{ By contrast, the transformations with $\varepsilon \not= 0$ but $\overline{\epsilon} = \overline{\zeta}^{r} = 0$ are proper gauge transformations. Thus,
the symmetry turns out to be two-fold} (if $\alpha\neq0$), labelled
by the leading order parameters $\overline{\epsilon}$ and $ \overline{\zeta}^{r} $.

Using 
 the proper gauge transformations, one can force the field $\chi$ to reduce to its leading term,
\begin{equation}
\chi=\overline{\chi}\,,
\end{equation}
which justifies why it is called a surface field. 

It is curious that the bulk term of the gauge transformations with
$\epsilon=-\varepsilon$ identically vanishes. These are improper
when{ $\overline{\zeta}^{r}\neq0$}. Thus, there exist improper gauge transformations
that are generated by pure boundary terms. This is possible because
the (pre-)symplectic form involves non trivial surface contributions,
as it happens in the case of the duality-invariant formulation of
electromagnetism \cite{Henneaux:2020nxi}.

{We shall use both parametrizations $G_{\epsilon,\varepsilon,\sigma^{k}}^{\text{RS}}$ and $G_{\epsilon,\varepsilon,\zeta^{k}}^{\text{RS}}$ of the gauge generators.  The first one is the natural one when one starts from the Rarita-Schwinger formulation of the action, as we have done here. The second one is more adapted to the description of the independent improper gauge symmetries, since these are just characterized by the simple conditions $\overline{\epsilon} = 0$, $\overline{\zeta}^{r}=0$, which take a ``non-diagonal form'' that mixes the parameters in the other parametrization.}

\subsection{Asymptotic symmetry algebra}

The bracket algebra of the canonical generators of gauge and Poincar\'e
symmetries  reads
\begin{align}
\{P_{\xi_{1},\xi_{1}^{i}},P_{\xi_{2},\xi_{2}^{i}}\} & =P_{\hat{\xi},\hat{\xi}^{i}}\,,\\
\{P_{\xi,\xi^{i}},G_{\epsilon,\varepsilon,\sigma^{k}}^{\text{RS}}\} & =G_{\hat{\epsilon},\hat{\varepsilon},\hat{\sigma}^{k}}^{\text{RS}}\,,\\
\{G_{\tau_{1}}^{\text{RS}},G_{\tau_{2}}^{\text{RS}}\} & =\mathcal{C}_{\{\tau_{1},\tau_{2}\}}\,, \label{eq:GRSGRS_C}
\end{align}
where $\tau_{i}=\{\epsilon_{i},\varepsilon_{i},\sigma_{i}^{k}\}$
with $i=1,2$, and where the hatted parameters of the commutator transformations are given by
\begin{align}
\hat{\xi} & =\xi_{1}^{i}\partial_{i}\xi_{2}-\xi_{2}^{i}\partial_{i}\xi_{1}\,,\label{eq:hatBoosts}\\
\hat{\xi}^{i} & =\xi_{1}^{j}\partial_{j}\xi_{2}^{i}-\xi_{2}^{j}\partial_{j}\xi_{1}^{i}+\xi_{1}\partial^{i}\xi_{2}-\xi_{2}\partial^{i}\xi_{1}\,,\label{eq:hatRotations}\\
\hat{\epsilon} & ={-\xi \gamma_0 \gamma^m \partial_m \epsilon} + \frac{1}{2}\partial_{j}\xi\gamma^{j}\gamma_{0}\epsilon +\mathcal{L}_{\xi} \epsilon  \, \label{eq:hatEpsilon}\\
\hat{\varepsilon}& ={-\xi \gamma_0 \gamma^m \partial_m \varepsilon}+\frac{1}{2}\partial_{m}\xi\gamma^{m}\gamma_{0}\varepsilon+\mathcal{L}_{\xi}\varepsilon\,,\label{eq:hatVarepsilon}\\
\hat{\sigma}^r &{= -\gamma^r \overline{D}_A (b \overline{D}^A \varepsilon) -\frac{1}{2} b \overline{\gamma}_A\overline{D}^A\varepsilon +b\gamma_0 \overline{\gamma}^A \overline{D}_A {\sigma}^r + \frac{3}{2}b \gamma^r \gamma_0 \sigma^r -\frac{1}{2} \partial_A b \overline{\gamma}^A \gamma_0 \sigma^r    +\mathcal{L}_{Y}{\sigma}^{r}  }\, ,
\end{align}
Only the asymptotic parts $\overline{\hat {\tau}}$ of the gauge parameters -- the ones defining the improper transformations-- are actually meaningful since one has always the freedom of adding proper gauge transformations to $\hat{\tau}$.   This freedom can be frozen by fixing the gauge and using the Dirac bracket, but the meaningful, gauge invariant, asymptotic information on the algebra can already be extracted without going through this procedure.

The bracket algebra of the generators of the improper gauge transformations provides a projective representation of the algebra of the transformations themselves, with a central extension $\mathcal{C}_{\{\tau_{1},\tau_{2}\}}$ that can be non trivial \cite{Brown:1986nw,Brown:1986ed}.  Here,  non-trivial central charges are allowed because the fermionic improper gauge symmetries are abelian and indeed do occur.  They are given by
\begin{align}
\mathcal{C}_{\{\epsilon,\varepsilon\}} & =-\mathcal{C}_{\{\varepsilon,\epsilon\}}=\frac{i}{2}\alpha\oint d^{2}S_{k}\epsilon^{T}\gamma^{km}\partial_{m}\varepsilon\,,\label{eq:C1}\\
\mathcal{C}_{\{\epsilon,\sigma^{k}\}} & =-\mathcal{C}_{\{\sigma^{k},\epsilon\}}=-\frac{i}{2}\alpha\oint d^{2}S_{k}\epsilon^{T}\gamma_{0}\sigma^{k}\,.\label{eq:C2}
\end{align}
{A consistency check is that the central charge vanishes when one of the gauge transformations is a proper gauge symmetry, i.e., in the specific case considered here, $\zeta^k =  \sigma^k + \gamma_0 \gamma^{km} \partial_m \varepsilon =0$.}

We close this section with two observations.
\begin{itemize}
\item The leading orders $\overline{\epsilon}$, $\overline{\varepsilon}$, $\overline{\sigma}^r$ of the gauge parameters, characterizing the improper gauge transformations, are functions on the $2$-sphere transforming in infinite-dimensional  representations of the Lorentz group that can be read off from (\ref{eq:hatEpsilon})-(\ref{eq:hatSigma}).  (Covariantly) Constant spinor fields define a four-dimensional irreducible representation, the "zero mode" representation. 
\item To remove the zero mode of the fermionic improper gauge symmetry parametrized by $\overline{\epsilon}$, one could impose the transverse condition  $\partial_k \rho^k =0$ on $\rho^k$.  This is consistent with Lorentz invariance and makes the  zero mode of $\chi$ pure gauge.
\end{itemize}
More on this will be discussed elsewhere.


\section{Rigid supersymmetry: spin-$(1,\frac{3}{2})$ multiplet}
\label{sec:Spin132}

We shall now analyze the compatibility of the infinite-dimensional
asymptotic symmetries with the super-Poincar\'e algebra in four dimensions.
We start with the simplest case of the spin-$(1,\frac{3}{2})$ multiplet
under rigid supersymmetry, which corresponds to the sum of the free
Maxwell and free Rarita-Schwinger actions
\begin{equation}
S=S_{\text{EM}}+S_{\text{RS}}\,,\label{eq:S-EM-RS}
\end{equation}
where \cite{Henneaux:2018gfi}
\begin{align}
S_{\text{EM}} & =\int dt\Big\{\int d^{3}x\left(\pi^{i}\dot{A}_{i}+\pi_{\Psi}\dot{\Psi}\right)-\oint d^{2}S_{i}A^{i}\dot{\Psi}\\
 & \quad-\int d^{3}x\left(\frac{1}{2}\pi^{i}\pi_{i}+\frac{1}{4}F^{ij}F_{ij}+\lambda\pi_{\Psi}-A_{t}\partial_{i}\pi^{i}\right)\Big\}\,,
\end{align}
and the Rarita-Schwinger action $S_{\text{RS}}$ is given in \eqref{eq:TotalAction}.

Under rigid supersymmetry the fields transform as
\begin{align}
\delta_{\epsilon_{0}}A_{k} & =i\epsilon_{0}^{T}\gamma_{0}\partial_{k}\chi+i\epsilon_{0}^{T}\gamma_{0}\mu_{k}\,,\\
\delta_{\epsilon_{0}}\pi^{k} & =-i\epsilon^{klm}\epsilon_{0}^{T}\gamma_{0}\gamma_{5}\partial_{l}\mu_{m}\,,\\
\delta_{\epsilon_{0}}\mu_{k} & =\frac{1}{2}\gamma_{m}\gamma_{k}W^{m}\gamma_{0}\epsilon_{0}\,,\\
\delta_{\epsilon_{0}}\chi & =0\,,
\end{align}
where $W^{k}=\pi^{k}-\epsilon^{klm}\partial_{l}A_{m}\gamma_{5}$,
and $\epsilon_{0}$ is a constant spinor parameter. These transformations
leave the piece of the total action \eqref{eq:S-EM-RS} involving only the fields $A_k$, $\pi^k$, $\mu_k$ and  $\chi$ invariant up to non-vanishing boundary
terms. We shall remedy this problem by 
defining the transformation laws of the boundary fields $\Psi$ and
$\rho^{k}$ in $S$ so that the symplectic form and the action be invariant
under rigid supersymmetry. 

This will not be sufficient for fully solving the problem, and an extra step must be simultaneously taken.   As all rigid symmetries are defined up to a gauge symmetry, we have
the freedom to add a gauge transformation to the supersymmetry transformations - denote it by $\nu$ - which can depend on the fields. The  variation of the symplectic form under supersymmetry and the accompanying gauge transformation
reads
$d_{V}(\iota_{\epsilon_{0}}\Omega+\iota_{\nu}\Omega)$. 
It turns out that one cannot define the supersymmetry transformations of the boundary fields so that $d_{V}(\iota_{\epsilon_{0}}\Omega) $ vanishes by itself.  It is also necessary  to add  a $U(1)$ gauge transformation
with a field dependent parameter
\begin{equation}
\nu_{(\epsilon_{0})}\equiv\mathcal{F}^{T}\epsilon_{0}\,,\qquad(\text{\ensuremath{\Rightarrow}}\nu_{(\epsilon_{0})}=0\text{ if }\epsilon_{0}=0)\,,
\end{equation}
where $\mathcal{F}$ is a function of the fermionic fields specified
below.  Only then can one achieve
\begin{equation}
d_{V}(\iota_{\epsilon_{0}}\Omega+\iota_{\nu}\Omega) = 0 \,,
\end{equation}
a condition that is necessary for integrability of the supersymmetry charges.

\subsection{Invariance under rigid supersymmetry}

The symplectic form for the multiplet $(1,\frac{3}{2})$ is explicitly given
by
\begin{equation}
\Omega=\Omega_{\text{EM}}+\Omega_{\text{RS}}\,,\label{eq:Omega(1,3/2)}
\end{equation}
 where
\begin{align}
\Omega_{\text{EM}} & =\int d^{3}x\left(d_{V}\pi^{k}d_{V}A_{k}+d_{V}\pi_{\Psi}d_{V}\Psi\right)-\oint d^{2}S_{k}d_{V}A^{k}d_{V}\Psi\,,\\
\Omega_{\text{RS}} & =-i\int d^{3}xd_{V}\chi^{T}\gamma^{km}\partial_{k}d_{V}\mu_{m}+\frac{i}{2}\int d^{3}xd_{V}\mu_{k}^{T}\gamma^{km}d_{V}\mu_{m}\nonumber \\
 & \quad+\frac{i}{2}\alpha\oint d^{2}S_{k}\Big(d_{V}\chi^{T}\gamma^{km}d_{V}\mu_{m}-d_{V}\chi^{T}\gamma_{0}d_{V}\rho^{k}\Big)\,.\label{eq:Omega-RS}
\end{align}

Under rigid supersymmetry, it changes as
\begin{align}
d_{V}(\iota_{\epsilon_{0}}\Omega) & = {\int d^3x \left(  d_{V}\delta_{\epsilon_{0}}\pi_{\Psi}d_{V}\Psi+d_{V}\pi_{\Psi}d_{V}\delta_{\epsilon_{0}}\Psi  \right)} \nonumber \\
&{\quad +\oint d^{2}S_{m}\left(id_{V}\pi^{m}d_{V}\chi^{T}\gamma_{0}\epsilon_{0}-d_{V}A^{m}d_{V}\delta_{\epsilon_{0}}\Psi\right)}\nonumber \\
 & \quad-\alpha i\oint d^{2}S_{m}\Big(\frac{1}{2}d_{V}\chi^{T}d_{V}W^{m}\gamma_{0}\epsilon_{0}+d_{V}\chi^{T}\gamma_{0}d_{V}\delta_{\epsilon_{0}}\rho^{m}\Big)\,.
\end{align}
The second term can be made zero by imposing the following transformation
law for the boundary field $\rho^{m}$
\begin{equation}
\delta_{\epsilon_{0}}\rho^{m}=\frac{1}{2}\gamma_{0}W^{m}\gamma_{0}\epsilon_{0} \quad {\Leftrightarrow \quad \delta_{\epsilon_0}\omega^m=0}\, .
\end{equation}
The first term can be removed by applying the field dependent $U(1)$
gauge transformation
\begin{equation}
d_{V}(\iota_{\nu}\Omega)=i\oint d^{2}S_{m}d_{V}\pi^{m}d_{V}\nu\,,
\end{equation}
with $\nu=\chi^{T}\gamma_{0}\epsilon_{0}$, and choosing $\delta_{\epsilon_{0}}\pi_{\Psi}=\delta_{\epsilon_{0}}\Psi=0$.
We then obtain that
\begin{equation}
d_{V}(\iota_{\epsilon_{0}}\Omega+\iota_{\nu}\Omega)=0\,.
\end{equation}

\subsection{Canonical realization of the asymptotic symmetries}

Once supersymmetry is defined on the fields as above, the theory is invariant under the rigid super-Poincar\'e transformations.  It is also invariant under the improper gauge symmetries of the spin-$1$ and spin-$\frac32$ fields, which are unaffected by the construction.  We examine their generators in turn.

\subsubsection{Super-Poincar\'e algebra}

The canonical generators of Poincar\'e transformations are given by the sum
\begin{align}
P_{\text{\ensuremath{\xi,\xi^{i}}}} & =P_{\xi,\xi^{i}}^{\text{EM}}+P_{\xi,\xi^{i}}^{\text{RS}}\,,
\end{align}
since the spin-$1$ and spin-$\frac32$ fields are uncoupled.  The charges are given by 
\begin{align}
P_{\xi,\xi^{i}}^{\text{EM}} & =\int d^{3}x(\xi\mathcal{H}^{\text{\text{EM}}}+\xi^{i}\mathcal{H}_{i}^{\text{\text{EM}}})+\mathcal{B}_{\xi,\xi^{i}}^{\text{EM}}\,,\\
\mathcal{H}^{\text{\text{EM}}} & =-\Psi\partial_{i}\pi^{i}-A_{i}\partial^{i}\pi_{\Psi}+\frac{1}{2}\pi_{i}\pi^{i}+\frac{1}{4}F_{ij}F^{ij}\,,\\
\mathcal{H}_{i}^{\text{\text{EM}}} & =F_{ij}\pi^{j}-A_{i}\partial_{j}\pi^{j}+\pi_{\Psi}\partial_{i}\Psi\,,\\
\mathcal{B}_{\xi,\xi^{i}}^{\text{EM}} & =\oint d^{2}x\Big[b(\bar{\Psi}\bar{\pi}^{r}+\sqrt{\bar{\gamma}}\bar{A}_{B}\bar{D}^{B}A_{r})+Y^{B}(\bar{A}_{B}\bar{\pi}^{r}+\sqrt{\bar{\gamma}}\bar{\Psi}\partial_{B}\bar{A}_{r})\Big]\,.
\end{align}
for the electromagnetic field \cite{Henneaux:2018hdj,Henneaux:2018gfi}  and $P_{\xi,\xi^{i}}^{\text{RS}}$ given in \eqref{eq:Poincare-RS} for the Rarita-Schwinger field. 

Infinitesimal Poincar\'e transformations for Maxwell fields read \cite{Henneaux:2018hdj,Henneaux:2018gfi}
\begin{align}
\delta_{\xi}A_{i} & =\xi\pi_{i}+\mathcal{L}_{\xi}A_{i}\,,\\
\delta_{\xi}\pi^{i} & =\partial_{j}(\xi F^{ji})+\mathcal{L}_{\xi}\pi^{i}\,,\\
\delta_{\xi}\Psi & =\partial_{i}(\xi A^{i})+\xi^{i}\partial_{i}\Psi\,,\\
\delta_{\xi}\pi_{\Psi} & =\xi \partial_i \pi^i + \partial_i (\xi^i \pi_\Psi) \approx 0 \, ,
\end{align}
and the ones for Rarita-Schwinger fields are given in \eqref{eq:PoincareChi},
\eqref{eq:PoincareMu}, \eqref{eq:PoincareRho}.

The canonical generator of rigid supersymmetry follows from the relation $\iota_{\epsilon_{0}}\Omega+\iota_{\nu(\epsilon_0)}\Omega = - d_V G_{\epsilon_{0}}^{\text{\ensuremath{\text{SUSY}}}}$ and is given by
\begin{equation}
G_{\epsilon_{0}}^{\text{\ensuremath{\text{SUSY}}}}=i\int d^{3}x\mu_{k}^{T}W^{k}\gamma_{0}\epsilon_{0}\,.
\end{equation}
Infinitesimal rigid supersymmetry transformations of all the fields
in the multiplet are then
\begin{align}
\delta_{\epsilon_{0}}A_{k} & =i\epsilon_{0}^{T}\gamma_{0}\mu_{k}\,,\\
\delta_{\epsilon_{0}}\pi^{k} & =-i\epsilon^{klm}\epsilon_{0}^{T}\gamma_{0}\gamma_{5}\partial_{l}\mu_{m}\,,\\
\delta_{\epsilon_{0}}\mu_{k} & =\frac{1}{2}\gamma_{m}\gamma_{k}W^{m}\gamma_{0}\epsilon_{0}\,,\\
\delta_{\epsilon_{0}}\rho^{m} & =\frac{1}{2}\gamma_{0}W^{m}\gamma_{0}\epsilon_{0}\,,\\
\delta_{\epsilon_{0}}\Psi & =\delta_{\epsilon_{0}}\chi=0\,.
\end{align}

It is straightforward to verify that the super-Poincar\'e algebra is
satisfied
\begin{align}
\{P_{\xi_{1},\xi_{1}^{i}},P_{\xi_{2},\xi_{2}^{i}}\} & =P_{\hat{\xi},\hat{\xi}^{i}}\,,\\
\{P_{\xi,\xi^{i}},G_{\epsilon_{0}}^{\text{SUSY}}\} & =G_{\hat{\epsilon}_{0}}^{\text{SUSY}}\,,\\
\{G_{\epsilon_{0}^{1}}^{\text{SUSY}},G_{\epsilon_{0}^{2}}^{\text{SUSY}}\} & =P_{\hat{a},\hat{a}^{i}}\,,\label{eq:GsusyGsusy}
\end{align}
where the hatted parameters of the commutator transformation are
\begin{align}
\hat{\epsilon}_{0} & =\frac{1}{2}\partial_{m}\xi\gamma^{m}\gamma_{0}\epsilon_{0}+\mathcal{L}_{\xi}\epsilon_{0}\,,\\
\hat{a} & =\frac{i}{2}(\epsilon_{0}^{2})^{T}\epsilon_{0}^{1}\,,\qquad\hat{a}^{i}=-\frac{i}{2}(\epsilon_{0}^{2})^{T}\gamma^{i}\gamma_{0}\epsilon_{0}^{1}\,,
\end{align}
with $\hat{\xi}$ and $\hat{\xi}^{i}$ given in \eqref{eq:hatBoosts}
and \eqref{eq:hatRotations}, respectively. 

It is worth noting that the bracket between two rigid supersymmetry
generators \eqref{eq:GsusyGsusy} correctly closes to spatial and time translations,
as expected for the canonical realization of the super-Poincar\'e algebra
on the spin-$(1,3/2)$ multiplet.

\subsubsection{Infinite-dimensional  algebra of improper gauge symmetries}

Since the two fields in the supermultiplet do not interact, the symmetry sector of the improper gauge symmetries is just the direct sum of the asymptotic angle-dependent $u(1)$ transformations of the Maxwell field and the asymptotic angle-dependent fermionic transformations of the Rarita-Schwinger field.  The canonical generators of improper gauge symmetries are unchanged.  For completeness, we reproduce them here, together with their algebra, 
\begin{align}
G_{\mu,\nu}^{\text{\ensuremath{\text{EM}}}} & =\int d^{3}x(\mu\pi_{\Psi}-\nu\partial_{i}\pi^{i})+\oint d^{2}S_{m}(\nu\pi^{m}-\mu A^{m})\,,\\
G_{\epsilon,\varepsilon,\sigma^{k}}^{\text{RS}} & =i\int d^{3}x\left(\epsilon^{T}+\varepsilon^{T}\right)\gamma^{km}\partial_{k}\mu_{m}\nonumber \\
 & \quad+\frac{i}{2}\alpha\oint d^{2}S_{k}\left(\epsilon^{T}\gamma^{km}\mu_{m}-\chi^{T}\gamma^{km}\partial_{m}\varepsilon-\epsilon^{T}\gamma_{0}\rho^{k}+\chi^{T}\gamma_{0}\sigma^{k}\right)\,.
\end{align}
Transformation laws for the Maxwell field under gauge symmetries read
\begin{equation}
\delta_{\mu,\nu}\Psi=\mu\,,\qquad\delta_{\mu,\nu}A_{i}=\partial_{i}\nu\,,\qquad\delta_{\mu,\nu}\pi^{i}=\delta_{\mu,\nu}\pi_{\Psi}=0\,,
\end{equation}
while the ones for Rarita-Schwinger fields are given in \eqref{eq:GaugeChi},
\eqref{eq:GaugeMu}, \eqref{eq:GaugeRho}.

The brackets of improper gauge generators with Poincar\'e read
\begin{align}
\{P_{\xi,\xi^{i}},G_{\mu,\nu}^{\text{EM}}\} & =G_{\hat{\mu},\hat{\nu}}^{\text{EM}}\,,\\
\{P_{\xi,\xi^{i}},G_{\epsilon,\varepsilon,\sigma^{k}}^{\text{RS}}\} & =G_{\hat{\epsilon},\hat{\varepsilon},\hat{\sigma}^{k}}^{\text{RS}}\,,
\end{align}
where the parameters transform as
\begin{alignat}{2}
\hat{\nu} & =\xi\mu+\mathcal{L}_{\xi}\rho\,,\qquad & \hat{\mu} & =\partial^{m}(\xi\partial_{m}\nu)+\mathcal{L}_{\xi}\nu_{m}\,,
\end{alignat}
with $\hat{\epsilon}$, $\hat{\varepsilon}$ and $\hat{\sigma}$ given
in \eqref{eq:hatEpsilon}, \eqref{eq:hatVarepsilon} and \eqref{eq:hatSigma},
respectively.

The brackets between improper gauge symmetries form a centrally extended
Abelian gauge algebra
\begin{align}
\{G_{\mu_{1},\nu_{1}}^{\text{EM}},G_{\mu_{2},\nu_{2}}^{\text{EM}}\} & =0\,,\\
\{G_{\mu,\nu}^{\text{EM}},G_{\tau}^{\text{RS}}\} & =0\,,\\
\{G_{\tau_{1}}^{\text{\text{RS}}},G_{\tau_{2}}^{\text{\text{RS}}}\} & =\mathcal{C}_{\{\tau_{1},\tau_{2}\}}\,,
\end{align}
where $\tau_{i}=\{\epsilon_{i},\varepsilon_{i},\sigma_{i}^{k}\}$
with $i=1,2$, and $\mathcal{C}_{\{\tau_{1},\tau_{2}\}}$ are given
in \eqref{eq:C1} and \eqref{eq:C2}.


\section{Rigid supersymmetry: spin-$(2,\frac{3}{2})$ multiplet}
\label{sec:Spin232}

We now turn to the spin-$(2,\frac{3}{2})$ multiplet, which is the relevant one for analyzing the super-BMS extensions in four dimensions.  The procedure parallels the one we followed for the spin-$(1,\frac{3}{2})$ multiplet.

The system is described by the sum of the free
spin-2 (Pauli-Fierz theory) and free Rarita-Schwinger actions
\begin{equation}
S=S_{\text{PF}}+S_{\text{RS}}\,.\label{eq:S-PF-RS}
\end{equation}
The Hamiltonian action for the free spin-2 theory on a Minkowski background
reads
\begin{align}
S_{\text{PF}} & =\int dt\,d^{3}x\left(\pi^{ij}\dot{h}_{ij}-\mathcal{E}-n\mathcal{G}-n^{i}\mathcal{G}_{i}\right)\,,
\end{align}
where the energy density and the Hamiltonian constraints associated
to the Lagrange multipliers $n$ and $n^{i}$, are given by
\begin{align}
\mathcal{E} & =\pi^{ij}\pi_{ij}-\frac{\pi^{2}}{2}+\frac{1}{4}\partial_{k}h_{ij}\partial^{k}h^{ij}-\frac{1}{2}\partial_{j}h^{ij}\partial^{k}h_{ik}+\frac{1}{4}\partial_{i}h\partial^{i}h+\frac{1}{2}h\mathcal{G}\,,\\
\mathcal{G} & =\triangle h-\partial_{i}\partial_{j}h^{ij}\,,\\
\mathcal{G}_{i} & =-2\partial_{j}\pi_{i}^{j}\,,
\end{align}
respectively. This theory is left invariant by the following gauge
transformations
\begin{align}
\delta_{\zeta^{n}}h_{ij} & =\partial_{i}\zeta_{j}+\partial_{j}\zeta_{i}\,,\label{eq:deltah-spin2}\\
\delta_{\zeta}\pi^{ij} & =\partial^{i}\partial^{j}\zeta-\delta^{ij}\triangle\zeta\,.\label{eq:deltapi-spin2}
\end{align}
The Rarita-Schwinger action $S_{\text{RS}}$ is given in \eqref{eq:TotalAction}.

The piece of the total action \eqref{eq:S-PF-RS} containing only the fields $h_{mn}$, $\pi^{mn}$, $\mu_m$ and $\chi$ is invariant, up to a non-vanishing boundary
terms, under rigid supersymmetry transformations of the form 
\begin{align}
\delta_{\epsilon_{0}}h_{mn} & =\frac{i}{2}\epsilon_{0}^{T}\gamma_{0}\gamma_{(m}\partial_{n)}\chi+\frac{i}{2}\epsilon_{0}^{T}\gamma_{0}\gamma_{(m}\mu_{n)}\,,\\
\delta_{\epsilon_{0}}\pi^{mn} & =\frac{i}{4}\epsilon_{0}^{T}(\partial^{m}\partial^{n}\chi-\delta^{mn}\Delta\chi)\nonumber \\
 & \quad+\frac{i}{4}\epsilon_{0}^{T}\partial^{(m}\mu^{n)}-\frac{i}{4}\delta^{mn}\epsilon_{0}^{T}\partial_{k}\mu^{k}+\frac{i}{4}\epsilon_{0}^{T}\gamma_{0}\gamma_{5}\gamma^{(m}\epsilon^{n)pq}\partial_{p}\mu_{q}\,,\\
\delta_{\epsilon_{0}}\mu_{p} & =\frac{1}{4}\partial_{m}h_{np}\gamma^{mn}\epsilon_{0}+\frac{1}{2}K_{mp}\gamma_{0}\gamma^{m}\epsilon_{0}\,,
\end{align}
and $\delta_{\epsilon_{0}}\chi=0$, where $\epsilon_{0}$ is a constant
spinor parameter. In particular, the kinetic term is invariant but only up to a non-vanishing surface term.

The solution
to the problem of achieving strict invariance  is reached again by suitably choosing the transformation
law under rigid supersymmetry for the boundary field $\rho^{k}$ in
$S_{\text{RS}}$, and by performing appropriate field dependent spin-2
gauge transformations with parameters
\begin{equation}
\zeta_{(\epsilon_{0})}\equiv i\epsilon_{0}^{T}\mathcal{F}\,,\qquad\zeta_{(\epsilon_{0})}^{n}\equiv i\epsilon_{0}^{T}\mathcal{F}^{n}\,,\qquad(\text{\ensuremath{\Rightarrow}}\zeta_{(\epsilon_{0})}=\zeta_{(\epsilon_{0})}^{n}=0\text{ if }\epsilon_{0}=0)\,,
\end{equation}
where $\mathcal{F}$ and $\mathcal{F}^{n}$ are functions of the fermionic
dynamical fields, whose specific form is given in the next subsection.

\subsection{Invariance under rigid supersymmetry}

The symplectic form for the multiplet $(2,\frac{3}{2})$ reads as
follows
\begin{equation}
\Omega=\Omega_{\text{PF}}+\Omega_{\text{RS}}\,,\label{eq:Omega(2,3/2)}
\end{equation}
 where
\begin{align}
\Omega_{\text{PF}} & =\int d^{3}xd_{V}\pi^{mn}d_{V}h_{mn}\,,
\end{align}
and $\Omega_{\text{RS}}$ is given in \eqref{eq:Omega-RS}.

The change in the symplectic form under rigid supersymmetry is given
by
\begin{align}
d_{V}(\iota_{\epsilon_{0}}\Omega) & =-\frac{i}{2}\oint d^{2}S_{m}\partial_{n}d_{V}\chi^{T}\epsilon_{0}(d_{V}h^{mn}-\delta^{mn}d_{V}h)-\frac{i}{2}\oint d^{2}S_{m}d_{V}\pi^{mn}d_{V}\chi^{T}\gamma_{0}\gamma_{n}\epsilon_{0}\nonumber \\
 & \quad+\frac{i}{2}\alpha\oint d^{2}S_{m}d_{V}\chi^{T}\gamma_{0}\Big(-\frac{1}{4}\gamma_{0}\gamma^{mn}\gamma^{pq}\partial_{p}d_{V}h_{nq}\epsilon_{0}+\frac{1}{2}d_{V}\pi^{mn}\gamma_{n}\epsilon_{0}+d_{V}\delta_{\epsilon_{0}}\rho^{m}\Big)\,.\label{eq:dOmega2}
\end{align}
The term proportional to $\alpha$ can be made zero, by imposing that
the boundary field $\rho^{m}$ transforms as 
\begin{equation}
\delta_{\epsilon_{0}}\rho^{m}=\frac{1}{4}\gamma_{0}\gamma^{mn}\gamma^{pq}\partial_{p}h_{nq}\epsilon_{0}-\frac{1}{2}\pi^{mn}\gamma_{n}\epsilon_{0}  \quad {\Leftrightarrow \quad \delta_{\epsilon_0}\omega^m=0}\, .
\end{equation}
The remaining terms in \eqref{eq:dOmega2} can be removed through
the following field dependent spin-2 gauge transformations
\begin{align}
d_{V}(\iota_{\zeta,\zeta^{k}}\Omega) & =\int d^{3}x\left(d_{V}\delta_{\zeta}\pi^{mn}d_{V}h_{mn}+d_{V}\pi^{mn}d_{V}\delta_{\zeta^{k}}h_{mn}\right)\nonumber \\
 & =2\oint d^{2}S_{m}\partial_{n}d_{V}\zeta(d_{V}h^{mn}-\delta^{mn}d_{V}h)+2\oint d^{2}S_{m}d_{V}\pi^{mn}d_{V}\zeta_{n}\,,
\end{align}
with parameters
\begin{alignat}{3}
\zeta_{(\epsilon_{0})} & =\frac{i}{4}\chi^{T}\epsilon_{0}\,, &  &  & \qquad\zeta_{(\epsilon_{0})}^{n} & =\frac{i}{4}\chi^{T}\gamma_{0}\gamma^{n}\epsilon_{0}\,.
\end{alignat}
We then obtain that the symplectic form is invariant
\begin{equation}
d_{V}(\iota_{\epsilon_{0}}\Omega+\iota_{\zeta,\zeta^{n}}\Omega)=0\,,
\end{equation}
which allows to define a canonical generator for rigid supersymmetry.

\subsection{Canonical realization of the asymptotic symmetries}

\subsubsection{Super-BMS algebra of \cite{Awada:1985by,Henneaux:2020ekh}}

The Poincar\'e canonical generators are again just the sum of the individual Poincar\'e generators and read
\begin{align}
P_{\text{\ensuremath{\xi,\xi^{i}}}} & =P_{\xi,\xi^{i}}^{\text{PF}}+P_{\xi,\xi^{i}}^{\text{RS}}\,,
\end{align}
where \cite{Fuentealba:2020ghw}
\begin{align}
P_{\xi,\xi^{i}}^{\text{PF}} & =\int d^{3}x(\xi\mathcal{H}^{\text{PF}}+\xi^{i}\mathcal{H}_{i}^{\text{PF}})+\mathcal{B}_{\xi,\xi^{i}}^{\text{PF}}\,,\\
\mathcal{H}^{\text{PF}} & =\pi^{ij}\pi_{ij}-\frac{\pi^{2}}{2}+\frac{1}{4}\partial_{k}h_{ij}\partial^{k}h^{ij}-\frac{1}{2}\partial_{j}h^{ij}\partial^{k}h_{ik}+\frac{1}{4}\partial_{i}h\partial^{i}h\nonumber \\
 & \quad+\partial_{l}(-h^{ij}\partial^{l}h_{ij}-h^{il}\partial_{i}h+\frac{3}{2}h^{lj}\partial^{i}h_{ij}+\frac{1}{2}h_{ij}\partial^{i}h^{jl})-\frac{1}{2}h(\partial_{i}\partial_{j}h^{ij}-\triangle h)\\
\mathcal{H}_{i}^{\text{PF}} & =-2\partial_{j}(\pi^{jk}h_{ik})+\pi^{jk}\partial_{i}h_{jk}\,,\\
\mathcal{B}_{\xi,\xi^{i}}^{\text{PF}} & =\oint d^{2}x\Big\{ b\Big[\sqrt{\bar{\gamma}}(-\frac{1}{2}\bar{h}\bar{h}_{rr}+\frac{1}{4}\bar{h}^{2}-\frac{3}{4}\bar{h}_{AB}\bar{h}^{AB})+\frac{2}{\sqrt{\bar{\gamma}}}\bar{\pi}_{A}^{r}\bar{\pi}^{rA}\Big]+2Y_{A}\bar{\pi}^{rB}\bar{h}_{B}^{A}\Big\}\,,
\end{align}
and $P_{\xi,\xi^{i}}^{\text{RS}}$ is written in \eqref{eq:Poincare-RS}.
Infinitesimal Poincar\'e transformations for Pauli-Fierz fields are
given by
\begin{align}
\delta_{\xi}h_{ij} & =2\xi(\pi_{ij}-\frac{1}{2}\delta_{ij}\pi)+\mathcal{L}_{\xi}h_{ij}\,,\\
\delta_{\xi}\pi^{ij} & =\frac{1}{2}\xi(\triangle h^{ij}+\partial^{i}\partial^{j}h-2\partial_{k}\partial^{(i}h^{j)k})\\
 & \quad\frac{1}{2}\partial_{k}\xi\left[\partial^{k}h^{ij}-2\partial^{(i}h^{j)k}+\delta^{ij}(2\partial_{l}h^{kl}-\partial^{k}h)\right]\\
 & \quad-\frac{1}{2}\delta^{ij}\xi(\triangle h-\partial_{i}\partial_{j}h^{ij})+\mathcal{L}_{\xi}\pi^{ij}\,,
\end{align}
while the ones for Rarita-Schwinger fields are given in \eqref{eq:PoincareChi},
\eqref{eq:PoincareMu} and \eqref{eq:PoincareRho}.  

The rigid supersymmetry canonical generator $G_{\epsilon_{0}}^{\text{\ensuremath{\text{SUSY}}}}$, determined through $\iota_{\epsilon_{0}}\Omega+\iota_{\zeta,\zeta^{n}}\Omega = - d_V G_{\epsilon_{0}}^{\text{\ensuremath{\text{SUSY}}}}$  turns out to be given
by
\begin{equation}
G_{\epsilon_{0}}^{\text{\ensuremath{\text{SUSY}}}}=i\int d^{3}x\Big[-\frac{1}{2}\mu_{m}^{T}\pi^{mn}\gamma_{0}\gamma_{n}\epsilon_{0}-\frac{1}{4}\mu_{m}^{T}(\partial_{n}h^{mn}-\partial^{m}h)\epsilon_{0}-\frac{1}{4}\epsilon_{0}^{T}\gamma^{i}\gamma^{jrs}\mu_{s}\partial_{r}h_{ij}\Big]\,.
\end{equation}
It reproduces through the Poisson bracket  the above supersymmetry transformation laws for the fields accompanied by the above field-dependent spin-2 gauge transformations, i.e., explicitly,
\begin{align}
\delta_{\epsilon_{0}}h_{mn} & =\frac{i}{2}\epsilon_{0}^{T}\gamma_{0}\gamma_{(m}\mu_{n)}\,,\\
\delta_{\epsilon_{0}}\pi^{mn} & =\frac{i}{4}\epsilon_{0}^{T}\partial^{(m}\mu^{n)}-\frac{i}{4}\delta^{mn}\epsilon_{0}^{T}\partial_{k}\mu^{k}+\frac{i}{4}\epsilon_{0}^{T}\gamma_{0}\gamma_{5}\gamma^{(m}\epsilon^{n)pq}\partial_{p}\mu_{q}\,,\\
\delta_{\epsilon_{0}}\mu_{p} & =-\frac{1}{4}\partial_{m}h_{np}\gamma^{mn}\epsilon_{0}-\frac{1}{2}K_{mp}\gamma_{0}\gamma^{m}\epsilon_{0}\,,\\
\delta_{\epsilon_{0}}\rho^{m} & =\frac{1}{4}\gamma_{0}\gamma^{mn}\gamma^{pq}\partial_{p}h_{nq}\epsilon_{0}-\frac{1}{2}\pi^{mn}\gamma_{n}\epsilon_{0}\,,
\end{align}
and $\delta_{\epsilon_{0}}\chi=0$. 

The canonical generator of the bosonic improper gauge transformations, which are the proper BMS supertranslations in the free theory \cite{Fuentealba:2020ghw},
reads
\begin{equation}
G_{T,W}=\int d^{3}x\Big[\zeta(\triangle h-\partial_{i}\partial_{j}h^{ij})-2\zeta^{i}\partial_{j}\pi_{i}^{j}\Big]+2\oint d^{2}x\Big[\sqrt{\bar{\gamma}}T\bar{h}_{rr}+W(\bar{\pi}^{rr}-\bar{\pi}_{A}^{A})\Big]\,.
\end{equation}
Infinitesimal gauge transformations are then given by
\begin{align}
\delta_{\zeta}h_{ij} & =\partial_{i}\zeta_{j}+\partial_{j}\zeta_{i}\,,\\
\delta_{\zeta}\pi^{ij} & =\partial^{i}\partial^{j}\zeta-\delta^{ij}\triangle\zeta\,,
\end{align}
with
\begin{alignat}{3}
\zeta= & T+\mathcal{O}\Big(\frac{1}{r}\Big)\,, &  &  & \qquad\zeta^{i} & =\partial^{i}(rW)+\mathcal{O}\Big(\frac{1}{r}\Big)\,,
\end{alignat}
where the set of functions $\{T^{\text{even}},W^{\text{odd}}\}$ turns
out to generate BMS supertranslations. The remaining parts $\{T^{\text{odd}},W^{\text{even}}\}$
drop out trivially of the {charge} due to the parity conditions
on the bosonic fields \cite{Henneaux:2018hdj,Henneaux:2019yax}, so these generate
proper gauge transformations.

One may then easily derive the Poisson bracket algebra of the generators of the super-Poincar\'e and BMS supertranslation generators, to get
\begin{align}
\{P_{\xi_{1},\xi_{1}^{i}},P_{\xi_{2},\xi_{2}^{i}}\} & =P_{\hat{\xi},\hat{\xi}^{i}}\,,\\
\{P_{\xi,\xi^{i}},G_{T,W}\} & =G_{\hat{T},\hat{W}}\,,\\
\{G_{T_{1},W_{1}},G_{T_{2},W_{2}}\} & =0\,,\\
\{P_{\xi,\xi^{i}},G_{\epsilon_{0}}^{\text{SUSY}}\} & =G_{\hat{\epsilon}_{0}}^{\text{SUSY}}\,,\\
\{G_{\epsilon_{0}^{1}}^{\text{SUSY}},G_{\epsilon_{0}^{2}}^{\text{SUSY}}\} & =P_{\hat{a},\hat{a}^{i}}\,,
\end{align}
where the parameters transform as
\begin{alignat}{1}
\hat{T} & =-3bW-\partial_{A}b\bar{D}^{A}W-b\bar{D}_{A}\bar{D}^{A}W+Y^{A}\partial_{A}T\,,\\
\hat{W} & =-bT+Y^{A}\partial_{A}W\,,\\
\hat{\epsilon}_{0} & =\frac{1}{2}\partial_{m}\xi\gamma^{m}\gamma_{0}\epsilon_{0}+\mathcal{L}_{\xi}\epsilon_{0}\,,\\
\hat{a} & =-\frac{i}{4}(\epsilon_{0}^{2})^{T}\epsilon_{0}^{1}\,,\qquad\qquad\qquad\hat{a}^{i}=-\frac{i}{4}(\epsilon_{0}^{2})^{T}\gamma_{0}\gamma^{i}\epsilon_{0}^{1}\,,
\end{alignat}
with $\hat{\xi}$ and $\hat{\xi}^{i}$ given in \eqref{eq:hatBoosts}
and \eqref{eq:hatRotations}, respectively.

This is just the super-BMS algebra of \cite{Awada:1985by,Henneaux:2020ekh}, with the finite number of fermionic generators $G_{\epsilon_{0}}^{\text{SUSY}}$ parametrized by a constant spinor (``restricted super-BMS algebra'').    To make the identification of the algebras, one must recall how the Poincar\'e translations and the improper spin-$2$ gauge symmetries of the linear theory combine to form the full set of BMS super-translations \cite{Fuentealba:2020ghw}.

\subsubsection{Infinite-dimensional fermionic gauge algebra}

But, as we have shown above,  there are more fermionic symmetries, which take the form of  improper gauge transformations.  These asymptotic symmetries are clearly unaffected by the inclusion of the free spin-2- field.  The full algebra, containing also these fermionic symmetries, is a graded extension of the BMS algebra with an infinite number of fermionic generators. 

The canonical generator of the improper fermionic gauge symmetries was worked out above,
\begin{align}
G_{\epsilon,\varepsilon,\sigma^{k}}^{\text{RS}} & =i\int d^{3}x\left(\epsilon^{T}+\varepsilon^{T}\right)\gamma^{km}\partial_{k}\mu_{m}\\
 & \quad + \frac{i}{2}\alpha\oint d^{2}S_{k}\left(\epsilon^{T}\gamma^{km}\mu_{m}-\chi^{T}\gamma^{km}\partial_{m}\varepsilon-\epsilon^{T}\gamma_{0}\rho^{k}+\chi^{T}\gamma_{0}\sigma^{{k}}\right)\,.
\end{align}
The brackets of this generator with the restricted super-BMS generators read
\begin{align}
\{P_{\xi,\xi^{i}},G_{\epsilon,\varepsilon,\sigma^{k}}^{\text{RS}}\} & =G_{\hat{\epsilon},\hat{\varepsilon},\hat{\sigma}^{k}}^{\text{\text{RS}}}\,,\\
\{G_{T,W},G_{\epsilon,\varepsilon,\sigma^{k}}^{\text{RS}}\} & =0\,,\\
\{G_{\epsilon_{0}}^{\text{SUSY}},G_{\epsilon,\varepsilon,\sigma^{k}}^{\text{RS}}\} & =0\,,
\end{align}
where $\hat{\epsilon}$, $\hat{\varepsilon}$ and $\hat{\sigma}$
are given in \eqref{eq:hatEpsilon}, \eqref{eq:hatVarepsilon} and
\eqref{eq:hatSigma}, respectively. The brackets between the improper
fermionic gauge symmetries form a centrally extended Abelian gauge
algebra, written in \eqref{eq:GRSGRS_C}.


\section{Conclusions}
\label{sec:Conclusions}

In this paper, we have consistently relaxed the boundary conditions at spatial infinity of the Rarita-Schwinger field in such a way that the resulting fermionic improper gauge symmetries form an infinite-dimensional algebra parametrized by two independent functions of the angles. Poincar\'e invariance is maintained in the sense that not only the boundary conditions are Poincar\'e invariant but also the action itself so that the Poincar\'e transformations have well-defined (integrable and finite) generators.  To achieve this result, we introduced boundary degrees of freedom at infinity, which modify the symplectic structure by boundary terms. We have also shown that the analysis can be extended to cover the supersymmetric free $(1,3/2)$ and $(2, 3/2)$ multiplets and that it is compatible with supersymmetry.  In the $(2, 3/2)$ case, one finds a symmetry superalgebra which is graded extension of the BMS algebra with an infinite number of fermionic generators.

It remains to be explored whether similar results still hold when interactions are switched on, i.e. in supergravity.  As the Yang-Mills example shows, this is not automatic \cite{Tanzi:2020fmt}.  Work along these lines is in progress. 

\section*{Acknowledgments}

O.~F. holds a ``Marina Solvay'' fellowship. This work was partially supported by the ERC Advanced Grant ``High-Spin-Grav'',  by FNRS-Belgium (conventions FRFC PDRT.1025.14 and  IISN 4.4503.15), as well as by funds from the Solvay Family.


\appendix

\section{Notations and conventions \label{A}}

Spinor fields will be mostly considered in two different orthonormal frames.  One is the ``cartesian frame'' formed by the vectors $\{\frac{\partial}{\partial x^i}\}$.   The other is the ``spherical frame'' given by $\{{e_{(a)}}^{k}\frac{\partial}{\partial x^{k}}\equiv\frac{\partial}{\partial r},\frac{1}{r}\frac{\partial}{\partial\theta},\frac{1}{r\sin\theta}\frac{\partial}{\partial\varphi}\}$ in spherical coordinates. 

When dealing with the cartesian frame, it is natural to use cartesian coordinates.  The associated Christoffel symbols and spin connection clearly vanish.

We give the corresponding formulas in the spherical frame, and use spherical coordinates. Indices in parentheses refer to the local frames,
while indices without parentheses are coordinate indices. The matrix
$({e_{(a)}}^{k})$ is given in spherical coordinates by 
\begin{equation}
({e_{(a)}}^{k})=\begin{pmatrix}1 & 0 & 0\\
0 & \frac{1}{r} & 0\\
0 & 0 & \frac{1}{r\sin\theta}
\end{pmatrix}\,.
\end{equation}

The non-vanishing Christoffel symbols in spherical coordinates are
\begin{eqnarray}
 &  & {\Gamma^{\theta}}_{\theta r}={\Gamma^{\theta}}_{r\theta}=\frac{1}{r},\quad{\Gamma^{r}}_{\theta\theta}=-r\,,\\
 &  & {\Gamma^{\varphi}}_{\varphi r}={\Gamma^{\varphi}}_{r\varphi}=\frac{1}{r},\quad{\Gamma^{r}}_{\varphi\varphi}=-r\sin^{2}\theta\,,\\
 &  & {\Gamma^{\varphi}}_{\varphi\theta}={\Gamma^{\varphi}}_{\theta\varphi}=\frac{\cos\theta}{\sin\theta},\quad{\Gamma^{\theta}}_{\varphi\varphi}=-\cos\theta\sin\theta\,,
\end{eqnarray}
while the spin connection coefficients $\omega_{(a)(b)m}=-\omega_{(b)(a)m}$
defined as 
\begin{eqnarray}
\omega_{(a)(b)m} & = & e_{(a)k\vert m}e_{(b)}^{\quad k}\,,
\end{eqnarray}
fulfill the triad condition 
\begin{equation}
\nabla_{m}e_{(a)}^{\quad k}=e_{(a)\;\,\vert m}^{\quad k}-\omega_{(a)\;\,m}^{\;\;(b)}e_{(b)}^{\quad k}=0\,,
\end{equation}
where $e_{(a)\;\,\vert m}^{\quad k}=e_{(a)\;\,,m}^{\quad k}+{\Gamma^{k}}_{mn}e_{(a)}^{\quad n}$.
The spin connections coefficients are explicitly given in the spherical
frame by 
\begin{eqnarray}
 &  & \omega_{(a)(b)r}=0, \nonumber \\
 &  & \omega_{(1)(2)\theta}=1=-\omega_{(2)(1)\theta},\quad\omega_{(1)(3)\theta}=0=-\omega_{(3)(1)\theta},\quad\omega_{(2)(3)\theta}=0=-\omega_{(3)(2)\theta},\\
 &  & \omega_{(1)(2)\varphi}=0=-\omega_{(2)(1)\varphi},\quad\omega_{(1)(3)\varphi}=\sin\theta=-\omega_{(3)(1)\varphi},\quad\omega_{(2)(3)\varphi}=\cos\theta=-\omega_{(3)(2)\varphi}, \nonumber
\end{eqnarray}

\vspace{0.1cm}

The $\gamma$-matrices are 
\begin{eqnarray}
 &  & \gamma_{(0)}=\begin{pmatrix}0 & 1 & 0 & 0\\
-1 & 0 & 0 & 0\\
0 & 0 & 0 & -1\\
0 & 0 & 1 & 0
\end{pmatrix},\quad\gamma_{(1)}=\begin{pmatrix}1 & 0 & 0 & 0\\
0 & -1 & 0 & 0\\
0 & 0 & -1 & 0\\
0 & 0 & 0 & 1
\end{pmatrix}\\
 &  & \gamma_{(2)}=\begin{pmatrix}0 & 0 & -1 & 0\\
0 & 0 & 0 & -1\\
-1 & 0 & 0 & 0\\
0 & -1 & 0 & 0
\end{pmatrix}\quad\gamma_{(3)}=\begin{pmatrix}0 & -1 & 0 & 0\\
-1 & 0 & 0 & 0\\
0 & 0 & 0 & 1\\
0 & 0 & 1 & 0
\end{pmatrix}\,,
\end{eqnarray}
where $\gamma_{(r)}$ coincides in the spherical frame with $\gamma_{(1)}$.
The spin covariant derivative of the $\gamma$-matrices in spherical
coordinates $\gamma_{m}$, and in the local frame $\gamma_{\left(a\right)}$,
\begin{eqnarray}
\nabla_{k}\gamma_{m} & = & \partial_{k}\gamma_{k}-{\Gamma^{n}}_{km}\gamma_{n}-\frac{1}{4}\omega_{(c)(d)k}[\gamma^{(c)(d)},\gamma_{m}]=0\,,
\end{eqnarray}
\begin{equation}
\nabla_{k}\gamma_{(a)}=\partial_{k}\gamma_{(a)}-\frac{1}{4}\omega_{(c)(d)k}[\gamma^{(c)(d)},\gamma_{(a)}]-\omega_{(a)(b)k}\gamma^{(b)}=0\,,
\end{equation}
 vanish. The covariant derivative of spinor fields and vector-spinor
fields are respectively given by 
\begin{eqnarray}
\nabla_{k}\chi & = & \partial_{k}\chi-\frac{1}{4}\omega_{(a)(b)k}\gamma^{(a)(b)}\chi\,,\\
\nabla_{k}\mu_{m} & = & \partial_{k}\mu_{m}-\Gamma_{\,\,km}^{n}\mu_{n}-\frac{1}{4}\omega_{(a)(b)k}\gamma^{(a)(b)}\mu_{m\,.}
\end{eqnarray}

\section{Asymptotic conditions and charges in spherical coordinates} \label{Spherical}

In spherical coordinates, the radial and angular components of the
spinor fields fall-off as
\begin{equation}
\chi=\overline{\chi}+\mathcal{O}\left(\frac{1}{r}\right)\,,
\end{equation}

\begin{align}
\mu_{r}=\frac{\overline{\mu}_{r}}{r^{2}}+\mathcal{O}\left(\frac{1}{r^{3}}\right)\,, & \quad\mu_{A}=\frac{\overline{\mu}_{A}}{r}+\mathcal{O}\left(\frac{1}{r^{2}}\right)\,,\nonumber \\
\rho_{r}=\frac{\overline{\rho}_{r}}{r^{2}}+\mathcal{O}\left(\frac{1}{r^{3}}\right)\,, & \quad\rho_{A}=\frac{\overline{\rho}_{A}}{r}+\mathcal{O}\left(\frac{1}{r^{2}}\right)\,.
\end{align}
The Poincar\'e parameters behave as 
\begin{align}
\xi=rb+T\,,\quad\xi^{r}=W\,,\quad\xi^{A}=Y^{A}+\frac{1}{r}\overline{D}^{A}W\,,
\end{align}
where the asymptotic Killing equations read
\begin{equation}
\overline{D}^{A}\overline{D}^{B}W+\overline{\gamma}_{AB}W=0\,,\quad\overline{D}_{A}\overline{D}_{B}b+\overline{\gamma}_{AB}b=0\,,\quad\mathcal{L}_{Y}\overline{\gamma}_{AB}=0\,,\quad\partial_{A}T=0\,.
\end{equation}
The functions on the sphere, $b$ and $Y^{A}$ describe the homogeneous
Lorentz transformations, while $T$ and $W$ describe translations.
Here, $\overline{D}^{A}$ denotes the covariant derivative with respect
to the metric on the $2$-sphere. The gauge parameters behave as
\begin{eqnarray}
\epsilon & = & \overline{\epsilon}+\mathcal{O}\left(\frac{1}{r}\right)\,,\\
\varepsilon & = & \frac{\overline{\varepsilon}}{r}+\mathcal{O}\left(\frac{1}{r^{2}}\right)\,,\\
\sigma_{r} & = & \frac{\text{\ensuremath{\overline{\sigma}}}_{r}}{r^{2}}+\mathcal{O}\left(\frac{1}{r^{3}}\right)\,,\\
\sigma_{A} & = & \frac{\text{\ensuremath{\overline{\sigma}}}_{A}}{r}+\mathcal{O}\left(\frac{1}{r^{2}}\right)\,.
\end{eqnarray}

The explicit form of the constraint in spherical coordinates at leading
order reads 
\begin{eqnarray} \label{Constraint-LO}
\mathcal{S} & = & \frac{1}{{r}}\left(\overline{\gamma}^{AB}\overline{\nabla}_{A}\overline{\mu}_{B} {-\frac{1}{2}\gamma_1 \overline{\gamma}^A \overline{\mu}_A}    -\gamma^{r}\overline{\gamma}^{A}\overline{\nabla}_{A}\overline{\mu}_{r}-\overline{\mu}_{r}\right)+\mathcal{O}\left(\frac{1}{r^{{2}}}\right)\,,
\end{eqnarray}
where $\overline{\nabla}_{A}$ stands for the spin-covariant derivative
on the $2$-sphere.

\subsection*{Poincar\'e and fermionic gauge charges in spherical coordinates}

Poincar\'e generators are given by
\begin{eqnarray}
P_{\text{\ensuremath{\xi},\ensuremath{\xi^{i}}}} & = & \int d^{3}x\left(\xi\mathcal{H^{\text{RS}}}+\xi^{i}\mathcal{H}_{i}^{\text{RS}}\right)+\mathcal{B}_{\xi,\xi^{i}}^{\text{RS}}\,,
\end{eqnarray}
where the bulk pieces of the generators can be read off from (\ref{eq:Poincare-RS}), and the asymptotic charges become 
\begin{eqnarray}
\mathcal{B}_{\xi,\xi^{i}}^{\text{RS}} & = &  \frac{i}{2}\alpha  \oint d^2x \left( \delta_{b} \overline{\chi}^T \gamma_0 \overline{\omega}^r + \mathcal{L}_{Y}\overline{\chi}^T \gamma_0 \overline{\omega}^r  \right) \,.
\end{eqnarray}

Infinitesimal transformation laws under Poincar\'e are determined by
\begin{eqnarray}
\delta\mathcal{\overline{\chi}} & = &{-b\gamma_0 \overline{\gamma}^A \overline{D}_A \overline{\chi}} + \frac{3}{2}b\gamma^{r}\gamma_{0}\overline{\chi}+\frac{1}{2}\overline{\nabla}_{B}b\overline{\gamma}^{B}\gamma_{0}\overline{\chi}+\mathcal{L}_{Y}\overline{\chi}\,,\\
\delta\overline{\mu}_{r} & = & b\gamma_{0}\gamma^{r}\left(\overline{\gamma}^{BC}\overline{\nabla}_{B}\overline{\mu}_{C}+\gamma^{r}\overline{\gamma}^{B}\overline{\mu}_{B}\right)+\frac{1}{2}b\gamma^{r}\gamma_{0}\overline{\mu}_{r}+\frac{1}{2}\overline{\nabla}_{B}b\overline{\gamma}^{B}\gamma_{0}\overline{\mu}_{r} \nonumber \\
 &  & +\frac{1}{2}b\gamma_{0}\gamma^{r}\left(\overline{\gamma}^{AB}\overline{\nabla}_{A}\overline{\mu}_{B}-\gamma^{r}\overline{\gamma}^{A}\overline{\nabla}_{A}\overline{\mu}_{r}-\overline{\mu}_{r}\right)+\mathcal{L}_{Y}\overline{\mu}_{r}\,,\\
\delta\overline{\mu}_{A} & = & b\gamma_{0}\overline{\gamma}_{A}^{\,\,BC}\overline{\nabla}_{B}\overline{\mu}_{C}+b\gamma_{0}\gamma^{r}\overline{\gamma}_{A}^{\,\,B}\left(\overline{\nabla}_{B}\overline{\mu}_{r}-\frac{1}{2}\gamma^{r}\overline{\gamma}_{B}\overline{\mu}_{r}-\frac{1}{2}\overline{\mu}_{B}\right) \nonumber \\
 &  & +\frac{1}{2}b\gamma_{0}\overline{\gamma}_{A}\left(\overline{\gamma}^{BC}\overline{\nabla}_{B}\overline{\mu}_{C}-\gamma^{r}\overline{\gamma}^{B}\overline{\nabla}_{B}\overline{\mu}_{r}-\overline{\mu}_{r}\right)+\mathcal{L}_{Y}\overline{\mu}_{A}\,,\\
\delta\overline{\rho}^r &=&-\gamma^r \overline{D}_A (b \overline{\mu}^A) -\frac{1}{2} b \overline{\gamma}_A\overline{\mu}^A +b\gamma_0 \overline{\gamma}^A \overline{D}_A \overline{\rho}^r + \frac{3}{2}b \gamma^r \gamma_0 \overline{\rho}^r \nonumber \\
& & -\frac{1}{2} \partial_A b \overline{\gamma}^A \gamma_0 \overline{\rho}^r    +\mathcal{L}_{Y}\overline{\rho}^{r}  \, , \\
\delta\overline{\rho}^{A} & = & 0 \,.
\end{eqnarray}
The surface charges of improper gauge symmetries are
\begin{eqnarray}
\mathcal{B}_{\epsilon,\varepsilon,\sigma^{k}}^{\text{RS}} & = & \frac{i}{2}\alpha\oint d^{2}x\left(\overline{\epsilon}^{T}\gamma^{r}\overline{\gamma}^{A}\overline{\mu}_{A}-\overline{\chi}^{T}\gamma^{r}\overline{\gamma}^{A}\overline{D}_{A}\overline{\varepsilon}-\overline{\epsilon}^{T}\gamma_{0}\overline{\rho}^{r}+\overline{\chi}^{T}\gamma_{0}\overline{\sigma}^{r}\right)\,,
\end{eqnarray}
where the infinitesimal transformation laws under gauge symmetries
at leading order are given by
\begin{equation}
\delta_{\epsilon}\mathcal{\overline{\chi}}=\overline{\epsilon}\,,\quad\delta_{\varepsilon}\overline{\mu}_{r}=-\overline{\varepsilon}\,,\quad\delta_{\varepsilon}\overline{\mu}_{A}=\partial_{A}\overline{\varepsilon}\,,\quad\delta_{\sigma}\overline{\rho}_{r}=\overline{\sigma}_{r}\,,\quad\delta_{\sigma}\overline{\rho}_{A}=\overline{\sigma}_{A}\,.
\end{equation}



\end{document}